\documentclass[twocolumn]{aastex631}
\usepackage{amsmath}
\usepackage{graphicx}

\begin{document}
	
	\title{Bridging Roche Lobe Overflow and micro-TDEs:\\ The Runaway Evolution of Eccentric Mass Transfer in Star-Black Hole Binaries}
	
	\author[0009-0004-4744-4597]{Tian-Shun Chen}
	\affiliation{Tsung-Dao Lee Institute, Shanghai Jiao-Tong University, 1 Lisuo Road, Shanghai 201210, China}
	\affiliation{Department of Physics, College of Sciences, Shanghai University, 99 Shangda Road, Shanghai 200444, China}
	
	\author[0000-0002-1934-6250]{Dong Lai}

	\affiliation{Tsung-Dao Lee Institute, Shanghai Jiao-Tong University, 1 Lisuo Road, Shanghai 201210, China}
	\affiliation{Department of Astronomy, Center for Astrophysics and Planetary Science, Cornell University, Ithaca, NY 14853, USA}

	\begin{abstract}
Binary systems may undergo mass transfer while maintaining significant
orbital eccentricities. Stellar-mass black holes (sBHs) can strip
stars on eccentric orbits and produce micro-tidal disruption events
(micro-TDEs). While previous hydrodynamical studies have focused on compact systems on the verge of disruption, the transition
between self-regulated eccentric mass transfer and runaway disruption
remains poorly understood. We present SPH simulations of a Sun-like
star interacting with a $10\,M_\odot$ sBH across a range of initial
eccentricities ($e_0=0.30$--$0.70$) and pericenter distances
($b_0=3.33$--$3.57$ in units of the tidal radius), tracking the systems
for tens to over 100 orbital periods. Our results reveal that
these binaries can evolve along two distinct pathways, dictated by the
competition between mass-transfer-driven stellar expansion and orbital
widening: (i) Runaway disruption ($b_0\lesssim 3.45$), in which mass
loss at pericenter drives adiabatic expansion of the stellar envelope,
leading to unstable Roche-lobe overflow and runaway disruption of the
star. The stripped debris forms a thick accretion flow with
hyper-Eddington accretion rates onto the sBH, potentially powering fast
X-ray/UV or blue/optical transients. (ii) Stable mass transfer
($b_0\gtrsim 3.57$), in which the binary settles into a long-lived,
stable mass-transfer phase lasting up to 150 orbits (the limit of our
simulation), regulated by orbital expansion from pericenter mass loss.
These eccentric mass-transfer events could manifest observationally as
repeating, quasi-periodic flares.
	\end{abstract}

	\section{Introduction} \label{sec:intro}

Mass transfer is a fundamental process governing the evolution of
binary stars and shaping the formation of diverse astrophysical
populations ranging from X-ray binaries to gravitational wave
progenitors. While Roche-lobe overflow (RLOF) in circular orbits has
been extensively studied, mass transfer in eccentric orbits remains a
theoretical challenge. Recent observations of post-interaction binaries
with significant eccentricities
\citep[e.g.,][]{2013A&A...559A..54V,2019CoSka..49..264V,2024MNRAS.529.3729S}
suggest that mass transfer often commences before tidal circularization
is complete. To address this, recent theoretical works have developed
semi-analytical frameworks to describe the secular orbital evolution of
eccentric binaries
\citep{2005MNRAS.358..544R,2007ApJ...667.1170S,2019ApJ...872..119H,2024OJAp....7E..48B,2009ApJ...702.1387S,2016ApJ...825...71D}.
For instance, \citet{2025arXiv250905243P} presented a model that
relaxes the classical assumption of instantaneous circularization and
allows for non-conservative mass transfer for a broad range of binary
mass ratios and eccentricities. However, such semi-analytical
approaches inevitably rely on simplified assumptions about the gas flow
associated with stellar mass loss. They typically employ orbit-averaged
equations and parameterized mass transfer rates based on the distance to
the Lagrangian points. Crucially, the transition from stable eccentric
mass loss/transfer to dynamical instability depends sensitively on the
competition between the adiabatic response of the star and the orbital
evolution \citep[e.g.,][]{2025arXiv250510611Y}. Capturing the breakdown of these
approximations requires full hydrodynamical treatment, especially where
the structural readjustment of the star becomes nonlinear.

Extensive hydrodynamical simulations have investigated close encounters
between stars and black holes (BHs) on parabolic or highly eccentric
orbits, which typically result in the complete or partial tidal
disruption of the star
\citep[e.g.,][]{2013ApJL...771L..28M,2015MNRAS.449..771G,2020ApJ...904..100R,2022ApJ...933..203K,2023ApJ...948...89K,2024A&A...685A..45V,2024ApJ...977...80C}.
While these studies span diverse stellar structures (ranging from simple
polytropes to realistic MESA models), BH masses (from stellar-mass to
supermassive), and pericenter distances (from mild tidal grazing to
deep, disruptive encounters), they almost exclusively focus on the
outcome of a single pericenter passage. In contrast, numerical
investigations of repeated, multi-orbital encounters remain largely
unexplored. Early work by \citet{2011ApJ...732...74G} simulated
repeating tidal encounters for a giant planet orbiting a star on an
initially eccentric ($e_0=0.9$) orbit with a mass ratio of $10^3$. More
recently, \citet{2025ApJ...979...40L} modeled multi-encounter scenarios
between a star and a supermassive black hole (mass ratio $10^6$) across
various eccentricities ($e_0=0.8,0.9$) and penetration factors
($\beta_0\equiv r_{\rm tide,0}/r_{p,0}=0.5,0.6,1$, where
$r_{\rm tide,0}$ and $r_{p,0}$ are the initial tidal radius and
pericenter distance, respectively; see Eq.~\ref{eq:tidal_radius}), finding that all
cases triggered runaway mass loss and full disruption within 3 to 10
orbits. For lower-eccentricity regimes ($e_0=0$--$0.6$),
\citet{2024ApJ...961..149X} simulated the first three encounters
between a $1$--$10\,M_\odot$ star and a $10\,M_\odot$ BH across a range
of pericenter distances ($b_0=\beta_0^{-1}=1$--$3$). They demonstrated that
the star can undergo a gradual, multi-orbital stripping process termed a
``tidal peeling event.'' In summary, existing simulations initialize
star-BH systems on already compact, deeply penetrating orbits, focusing
mainly on the terminal phase of the interaction.

On the other hand, semi-analytical studies of eccentric star-SMBH tidal
interactions
\citep[e.g.,][]{2013MNRAS.429.3040L,2021MNRAS.503..603L,2023ApJ...945...86L,2023MNRAS.524.6247L,2024MNRAS.527.4317L,2024ApJ...974...67L,2025arXiv250510611Y,yu2025dynamicaltidemodifiedroche}
suggest that these systems may follow different evolutionary tracks,
including stable mass transfer and unstable stellar disruption,
depending on the uncertain physics of tidal heating and the stellar
response to heating. The long-term dynamical pathway of
eccentric binaries, describing the transition from secular tidal
evolution to possible catastrophic peeling and disruption, remains
unexplored in a self-consistent hydrodynamical framework.

In this paper, we attempt to bridge the gap between semi-analytical
binary evolution models and violent tidal disruption simulations. We
investigate the long-term evolution of a Sun-like star ($1\,M_\odot$)
encountering a $10\,M_\odot$ stellar-mass black hole (sBH) using the
smoothed particle hydrodynamics (SPH) code \textsc{Phantom}
\citep{2018PASA...35...31P}. Unlike previous studies that start with
immediate disruption or fast mass loss, we initialize the system well
outside the tidal radius and follow it for tens to hundreds of orbital
periods. This setup allows us to capture the initial nonlinear tidal
evolution, the onset of eccentric mass transfer, and the subsequent
divergence between self-regulated and runaway outcomes. By varying the
initial binary pericenter distance, $b_0=\beta_0^{-1}$, and eccentricity
($e_0$), we examine how the competition between adiabatic stellar
expansion and orbital widening shapes the transition region. Our results
provide a unified physical picture connecting eccentric Roche-lobe
overflow to the onset of micro-TDEs and their possible ultraluminous
X-ray transient counterparts \citep{2025ApJ...994L..17O,2025arXiv250922779B}.

The rest of this paper is organized as follows. In Section~\ref{sec:methods},
we describe the SPH setup, the stellar model, the orbital initial
conditions, and the diagnostics used to identify bound stellar material
and orbital evolution. Section~\ref{sec:results} presents the fiducial
runaway case, Run A, and analyzes its morphology, mass loss history,
orbital response, instability mechanism, disk formation, accretion rate,
and numerical resolution test. Section~\ref{sec:results2} examines Run
B, a wider-pericenter case that settles into quasi-stable mass transfer
rather than runaway disruption. In Section~\ref{sec:threshold}, we use
Runs C--F to probe the threshold region between these two evolutionary
pathways and to assess the role of the initial eccentricity. We summarize
our findings and discuss their observational implications in
Section~\ref{sec:conclusions}.

	\section{METHODOLOGY AND SIMULATION SETUP} \label{sec:methods}
	
We use smoothed particle hydrodynamics (SPH) to simulate the interaction between a $1\,M_\odot$ main-sequence star and a black hole ($M_{\rm BH}=10\,M_\odot$), initially on an eccentric orbit.
	
	\subsection{Stellar Model and Initial Conditions}
	
	The initial stellar model is a zero-age main-sequence star (ZAMS) with $M_*=M_0=1\,M_\odot$ and $R_*=R_0=R_\odot$, generated using the one-dimensional stellar evolution code MESA \citep{2011ApJS..192....3P}. The internal density and pressure profiles from the MESA model are mapped onto a three-dimensional SPH particle distribution. The star is relaxed in isolation to ensure strict hydrostatic equilibrium before being placed on the binary orbit.

	The fundamental length scale characterizing the BH-star interaction is the tidal radius, $r_{\rm tide}$, defined as:
	\begin{equation}
		r_{\rm tide} = R_* \left( \frac{M_{\rm BH}}{M_*} \right)^{1/3}.
		\label{eq:tidal_radius}
	\end{equation}
	For eccentric orbits, the strength of tidal interaction is governed by the pericenter distance $r_p$ relative to $r_{\rm tide}$. While the Roche lobe concept is strictly defined only for circular orbits, we can use the instantaneous Roche lobe radius $R_L$ at pericenter, given by \citep{1971ARA&A...9..183P},
	\begin{equation}
		R_L(r_p) = 0.462 r_p \left( \frac{M_*}{M_* + M_{\rm BH}} \right)^{1/3},
		\label{eq:roche}
	\end{equation}
	to qualitatively evaluate the onset of mass loss/transfer.
	
	In this paper, we carry out simulations for several values of the initial binary eccentricity $e_0 = 0.3, 0.55, 0.7$ and the initial dimensionless pericenter distance $b_0 \equiv r_{p,0}/r_{\rm tide,0} = 3.33, 3.45, 3.57$ (see Table~\ref{tab:initial_conditions}), corresponding to $\beta_0=b_0^{-1} = 0.30, 0.29, 0.28$. Here and throughout the paper, the subscript ``0'' denotes the initial value. In each case, we initialize the system at the corresponding apocenter $r_{a,0} = r_{p,0}(1+e_0)/(1-e_0)$. The initial binary period is given by
	\begin{equation}
		P_0 = b_0^{3/2} \frac{2\pi}{(1-e_0)^{3/2}} \left(\frac{R_0^3}{GM_0}\right)^{1/2} \left(\frac{M_{\rm BH}}{M_{\rm BH}+M_0}\right)^{1/2}.
	\end{equation}
	As we will show, for $b_0=3.33$, mass transfer commences almost immediately, leading to the eventual disruption of the star, whereas for $b_0=3.57$, the system appears to reach a state of stable mass transfer.
	
\begin{table*}[htpb]
	\centering
	\caption{Parameters for hydrodynamical simulations}
	\label{tab:initial_conditions}
	\begin{tabular}{lccccc}
		\hline
		\hline
		Run & $e_0$ & $b_0$ & $P_0$ $\left[(R_0^3/GM_0)^{1/2},\,{\rm days}\right]$ & $t_{\rm disrupt}$ [$P_0$] \\
		\hline
		Run A & 0.55 & 3.33 & $120.6,\,2.22$ & $41^*$ \\
		Run B & 0.55 & 3.57 & $133.9,\,2.47$ & $150$ \\
		Run C & 0.55 & 3.45 & $127.2,\,2.34$ & $92^*$ \\
		Run D & 0.30 & 3.33 & $62.2,\,1.15$  & $32^*$ \\
		Run E & 0.70 & 3.33 & $221.5,\,4.08$ & $39^*$ \\
		Run F & 0.30 & 3.57 & $69.0,\,1.27$ & $35$ \\
		\hline
	\end{tabular}
	\tablecomments{The different columns give the initial eccentricity ($e_0$), dimensionless pericenter distance ($b_0$), and orbital period ($P_0$, in units of the initial stellar dynamical time and in days). The last column gives the approximate disruption time of the star (marked with $^*$) or the stop time of the simulation (unmarked).}
\end{table*}
	
\subsection{Code Configuration and Microphysics}

We use the SPH code \textsc{Phantom} \citep{2018PASA...35...31P} for our simulations. To capture shocks and prevent particle interpenetration, we employ the standard artificial viscosity implementation \citep{1983JCoPh..52..374M,1992ARA&A..30..543M}. The linear and quadratic artificial-viscosity coefficients are set to 1.0 and 2.0, respectively, providing the necessary dissipation to handle high-Mach number shocks while minimizing viscosity in shear flows.

We use an adiabatic equation of state,
\begin{equation}
	p = (\gamma - 1) \rho \epsilon,
\end{equation}
with $\gamma = 5/3$, where $p$ is pressure, $\rho$ is density, and $\epsilon$ is the specific internal energy. We employ $10^5$ SPH particles for our fiducial runs. To ensure numerical robustness, we also perform a convergence test with $2 \times 10^5$ particles (see Section~\ref{subsec:numerical_robustness}), which yields consistent results.

An important physical assumption for our calculation is the neglect of radiative cooling. The tidal interaction and mass transfer occur on the orbital timescale ($\sim$ days), which is orders of magnitude shorter than the thermal Kelvin-Helmholtz timescale ($\tau_{\rm KH} \sim 10^7$ years) for the whole star. If the dissipation associated with tides and mass loss occurs near the stellar surface, the thermal timescale $t_{\rm th}$ can be shorter than $\tau_{\rm KH}$, approximately by a factor $(\Delta r/R_*)^2$, where $\Delta r$ measures the distance of heat deposition from the stellar surface. In this paper, we consider the evolution of the star and binary system on a timescale much less than $t_{\rm th}$, so that the star does not have time to radiate away the excess energy generated by tidal dissipation and shocks. Thus, we explicitly disable radiative cooling while enabling shock heating. This ensures that the generated heat is trapped within the stellar envelope, driving the adiabatic expansion that is central to the runaway mechanism discussed in Section~\ref{subsec:instability_mech}.

Self-gravity is calculated using a binary tree algorithm. The time integration follows a leapfrog kick-drift-kick scheme, with the timestep controlled by the Courant-Friedrichs-Lewy (CFL) condition. In our simulations, the BH is treated as an absorbing sphere with sink radius ($r_{\rm sink}=500\,GM_{\rm BH}/c^2\simeq0.011\,R_0$).

	\subsection{Analysis of Mass Loss, Accretion, and Orbital Evolution}
	\label{Analysis}
	To quantitatively analyze the mass stripping process and the structural response of the star, as well as the orbital evolution during mass loss, we must accurately distinguish between the bound stellar material and the unbound debris. Since the star can experience significant tidal deformation and heating, a simple binding energy criterion based on kinetic energy and potential energy alone is insufficient. Instead, we utilize the Bernoulli parameter, which accounts for the internal enthalpy of the gas.
	
	\noindent$\bullet$ \textbf{Bound Mass Determination.} For each SPH particle $i$, the Bernoulli parameter $B_i$ is calculated relative to the star's center of mass:
	\begin{equation}
		B_i = \frac{1}{2} |\mathbf{v}_i - \mathbf{v}_{\rm com}|^2 + \Phi_i + h_i,
	\end{equation}
	where $\mathbf{v}_i$ is the particle velocity, $\mathbf{v}_{\rm com}$ is the velocity of the stellar center of mass, $\Phi_i$ is the gravitational potential calculated using the tree method, and $h_i = \epsilon_i + P_i/\rho_i=\gamma\epsilon_i$ is the specific enthalpy. Particles with $B_i < 0$ are considered to be hydrodynamically bound to the star.
	
\begin{figure}[!h]
		\centering
		\includegraphics[width=1\linewidth]{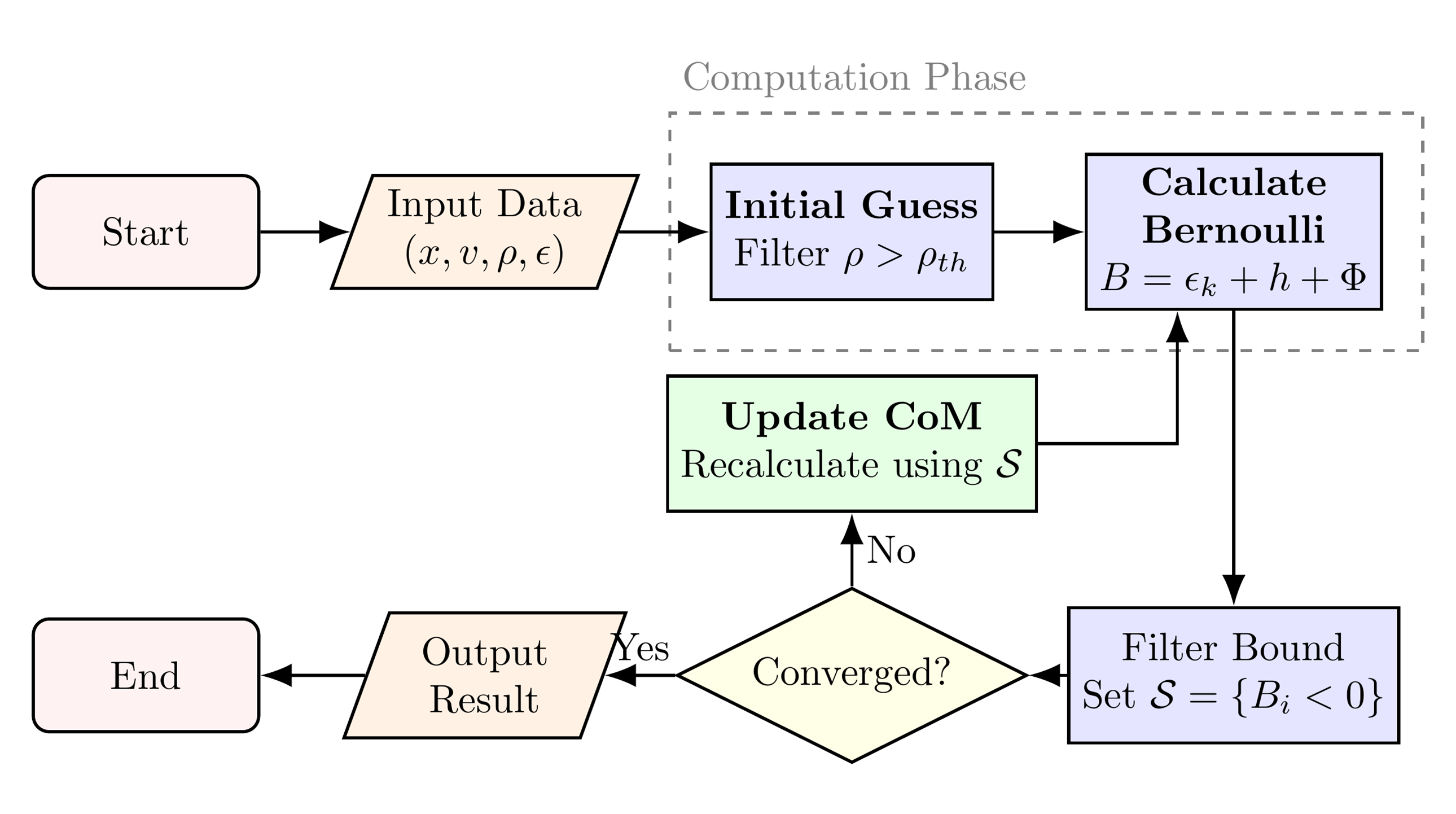}
		\caption{Iterative calculation of bound particles and stellar centroid. The initial centroid is determined by a 20\% maximum-density threshold; iteration is performed to update the bound particle set and centroid until convergence.}
		\label{fig:convergetest}
	\end{figure}

	Determining the bound mass and $\mathbf{v}_{\rm com}$ requires iteration. The initial iteration is obtained from the centroid of particles above 20\% of the maximum density. After identifying the first wave of bound particles based on $B_i<0$, a new center of mass is calculated using these particles. This iteration continues until the fractional difference between the bound particle sets of two consecutive iterations is less than $10^{-4}$, at which point the final bound particles and centroid are considered found. See Figure~\ref{fig:convergetest} for a flow chart of the iteration procedure.
	
	The mass of the star, $M_*(t)$, is the sum of the masses of all such bound particles. In our analysis, a Lagrangian radius is defined as the radial distance from the center of mass that encloses a specific fraction of the total bound mass. Accordingly, the effective radius of the star is characterized by $R_{90}$, which is the Lagrangian radius enclosing 90\% of the bound stellar mass.
	
	\noindent$\bullet$ \textbf{Orbit Determination.} To track the orbital evolution of the binary during the mass-stripping stage, we compute the semi-major axis and eccentricity of the bound star relative to the BH using two complementary methods.
	
	$(\mathrm{i})$ Dynamical method. We calculate the specific orbital energy $\mathcal{E}$ and specific angular momentum vector $\mathbf{h}$ of the star's center of mass:
	\begin{equation}
		\mathcal{E} = \frac{1}{2}v_{\rm rel}^2 - \frac{G(M_{\rm BH}+M_*)}{r_{\rm rel}}, \quad \mathbf{h} = \mathbf{r}_{\rm rel} \times \mathbf{v}_{\rm rel},
	\end{equation}
	where $\mathbf{r}_{\rm rel}$ and $\mathbf{v}_{\rm rel}$ are the position and velocity vectors relative to the BH. We evaluate these when the binary is around apocenter, where the orbital energy $\mathcal{E}$ is least affected by tidal effects. The semi-major axis and eccentricity vector are then given by
	\begin{equation}
		\begin{aligned}
		a_{\rm dyn} &= -\frac{G(M_{\rm BH}+M_*)}{2\mathcal{E}}, \\
		\mathbf{e}_{\rm dyn} &= \frac{\mathbf{v}_{\rm rel} \times \mathbf{h}}{G(M_{\rm BH}+M_*)} - \frac{\mathbf{r}_{\rm rel}}{r_{\rm rel}}.
		\end{aligned}
	\end{equation}
	The scalar eccentricity is $e_{\rm dyn} = |\mathbf{e}_{\rm dyn}|$.
	
	 $(\mathrm{ii})$ Geometric method. We compute the apocenter ($r_{a}$) and pericenter ($r_{p}$) distances between the stellar CoM and the BH, and obtain the semi-major axis and eccentricity from
	\begin{equation}
		a_{\rm geom} = \frac{r_{a}+r_{p}}{2}, \quad e_{\rm geom} = \frac{r_{a}-r_{p}}{r_{a}+r_{p}}.
	\end{equation}

	Because of the strong tidal interaction and mass loss around the pericenter, the orbit is not exactly Keplerian, and thus the orbital elements obtained using these two methods are not exactly the same.
	
	\noindent$\bullet$ \textbf{Accretion onto the BH.} Material is considered ``accreted'' onto the BH if it crosses the sink radius ($r_{\rm sink}=500\,GM_{\rm BH}/c^2\simeq0.011\,R_0$) of the BH. We track the cumulative accreted mass $M_{\rm acc}(t)$ and calculate the instantaneous mass accretion rate $\dot{M}_{\rm acc}$ by taking its time derivative.
	
	\section{Results and Analysis for Run A: \texorpdfstring{\lowercase{$b_0=3.33$, $e_0=0.55$}}{b0=3.33, e0=0.55}} \label{sec:results}
	
	In this section, we present the result for the simulation with $b_0=3.33$ (or $\beta_0=0.30$) and initial $e_0=0.55$. This case starts with tidal interaction and gentle mass loss, and ends with runaway disruption of the star.
	
	\subsection{Morphological Evolution}
	
	The transition from a detached binary to a violent runaway tidal disruption event is visually captured by the evolution of the gas morphology. Figure~\ref{fig:snapshots} displays a series of gas density slices taken in the orbital plane ($z=0$), illustrating four distinct phases of the star-BH interaction:
	
	\begin{figure*}[hbt!]
		\centering
		\includegraphics[width=1\textwidth]{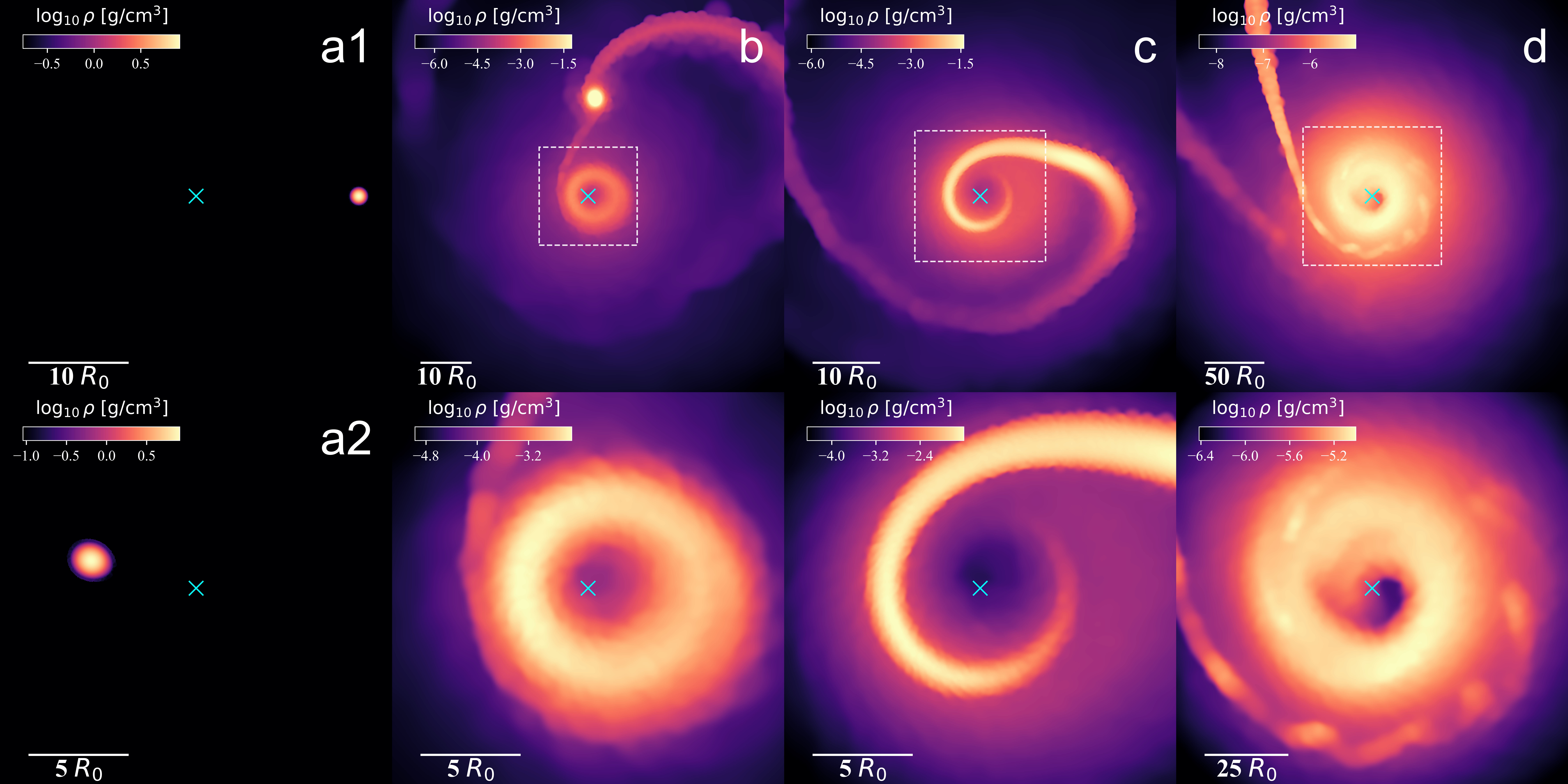} 
	\caption{Snapshots of gas density slices in the orbital plane ($z = 0$) for Run A (with $e_0=0.55$ and $b_0=3.33$). The cyan cross represents the location of the BH. For panels \textbf{(b)}, \textbf{(c)}, and \textbf{(d)}, the bottom sub-panels provide a zoomed-in view of the central region near the BH. The figure illustrates four distinct stages of the star-BH encounter: \textbf{(a1)} \& \textbf{(a2)} initial setup of the system at apocenter and the first pericenter passage ($t \approx 0.5 P_0$, with $P_0$ the initial orbital period), marking the onset of tidal interaction; \textbf{(b)} tidal peeling and disk formation at $t = 38 P_0$, showing the large-scale tidal tail and the circularizing eccentric disk; \textbf{(c)} terminal stellar disruption at $t = 41 P_0$, capturing the catastrophic shredding of the star into a long, coherent stream; and \textbf{(d)} debris fallback and disk expansion at $t = 60 P_0$, presenting the formation of a geometrically thick, shock-heated accretion torus.}
		\label{fig:snapshots}
	\end{figure*}
	
	\begin{enumerate}
		
		\item \textbf{Tidal Interaction ($t \lesssim 15 P_0$):} Panels (a1) and (a2) show the star at its initial configuration at the apocenter and around the first pericenter passage, respectively. In this early stage, the star acquires a teardrop shape at pericenter. It barely fills its Roche lobe, and mass loss is minimal as tidal dissipation drives the orbital decay.
		
		\item \textbf{Tidal Peeling and Disk Formation ($15 P_0\lesssim t \lesssim 40 P_0$):} As the orbit shrinks significantly, tidal interaction intensifies into the ``tidal peeling'' regime. Panels (b) reveal the formation of a substantial tidal tail wrapping around the BH (top). The zoom-in view (bottom) shows the stripped debris beginning to circularize and form an eccentric accretion disk around the BH.
		
		\item \textbf{Terminal Disruption ($t \simeq 41 P_0$):} The system enters the runaway phase. Panels (c) capture the catastrophic shredding of the stellar core, which is stretched into a long, coherent gas stream, marking the transition to complete disruption.
		
		\item \textbf{Debris Fallback and Disk Expansion ($t \gtrsim 41 P_0$):} Following the core disruption, the bound stellar debris falls back and shock-heats. Panels (d) show the system in the late stage, where the accretion disk has significantly expanded and thickened (see also Fig.~\ref{fig:disk_geometry} in Section~\ref{subsec:disk_structure}).
		
	\end{enumerate}
	
	\begin{figure*}[ht!]
		\centering
		\includegraphics[width=1\textwidth]{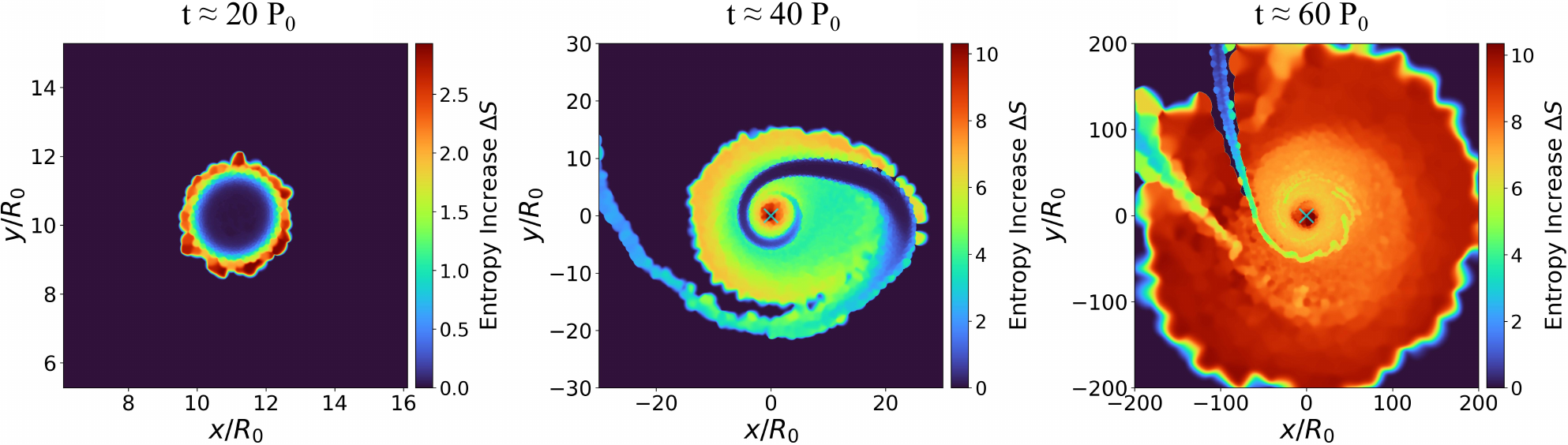}
		\caption{Spatial evolution of the entropy increase, $\Delta S = \ln(K/K_0)$ where $K = p/\rho^{5/3}$ and $K_0$ is the initial value, for Run A (see also Fig.~\ref{fig:snapshots}). The color map shows the thermodynamic state of the gas, with dark colors representing the gas that has experienced adiabatic evolution ($\Delta S \approx 0$) and bright colors indicating strong shock heating ($\Delta S \gtrsim 0$). The cyan cross shows the location of the BH. \textbf{Left} ($t \approx 20 P_0$, zoomed-in view of the star): During the early stage of mass transfer, the bulk of the star is largely adiabatic, and the envelope experiences shock heating. \textbf{Middle} ($t \approx 40 P_0$): The star is catastrophically disrupted into a tidal stream. Strong spiral shocks form at the stream self-intersection points, generating significant entropy. \textbf{Right} ($t \approx 60 P_0$): A geometrically thick, shock-heated accretion torus forms around the BH, thermodynamically distinct from the cool, adiabatically expanding tidal tail.}
		\label{fig:entropy_evolution}
	\end{figure*}
	
	To further elucidate the thermodynamics of the stellar material and the fate of the stellar debris, we map the spatial distribution of the gas entropy. Figure~\ref{fig:entropy_evolution} displays the evolution of the entropy increase, $\Delta S \equiv \ln(K/K_0)$, in the orbital plane at three representative epochs. Here, $K \equiv p/\rho^\gamma$, and $K_0$ is the initial value. Since an adiabatic flow conserves entropy ($\Delta S = 0$) while shocks irreversibly generate it ($\Delta S >~ 0$), the spatial evolution reveals a dramatic transition from an isentropic star to a shocked accretion flow.
	During the early mass transfer phase ($t \simeq 20 P_0$, Left Panel), the star remains largely intact and isentropic. The core exhibits $\Delta S \approx 0$, with minor entropy generation confined to the stellar surface where tidal stripping occurs. This state evolves into a catastrophic disruption around $t \approx 40 P_0$ (Middle Panel), capturing the system mid-runaway as the star is shredded into an elongated tidal stream. As the stream wraps around the BH and collides with itself, self-intersection shocks emerge as bright, high-entropy filaments (yellow/red regions). Following the disruption ($t \approx 60 P_0$, Right Panel), the bound stellar debris circularizes and settles into a geometrically thick accretion torus characterized by high entropy ($\Delta S \gg 1$), indicating a pressure-supported structure heated by shocks. In contrast, the unbound portion of the tidal stream---visible as a faint, low-entropy arc extending outward---continues to expand adiabatically, demonstrating a clear thermodynamic bifurcation between the bound, shock-heated disk and the unbound, ``pristine'' tidal tail.
	
	\subsection{Stellar Response: Adiabatic Expansion and Mass Loss History}
	\label{sec:stellar_response}
	The stability of mass transfer is fundamentally determined by how the stellar radius and internal structure respond to mass loss. For a star with a deep convective envelope, the material is effectively isentropic. Under the assumption of adiabatic mass loss (where the mass loss timescale is shorter than the thermal timescale), the stellar structure evolves according to the adiabatic mass-radius relation. For a polytrope with index $\gamma=5/3$, the radius scales as $R_* \propto M_*^{-1/3}$ \citep[e.g.,][]{1987ApJ...318..794H}. Consequently, a reduction in mass leads to a global \textit{expansion} of the star.
	
\begin{figure}[ht!]
		\centering    \includegraphics[width=1\linewidth]{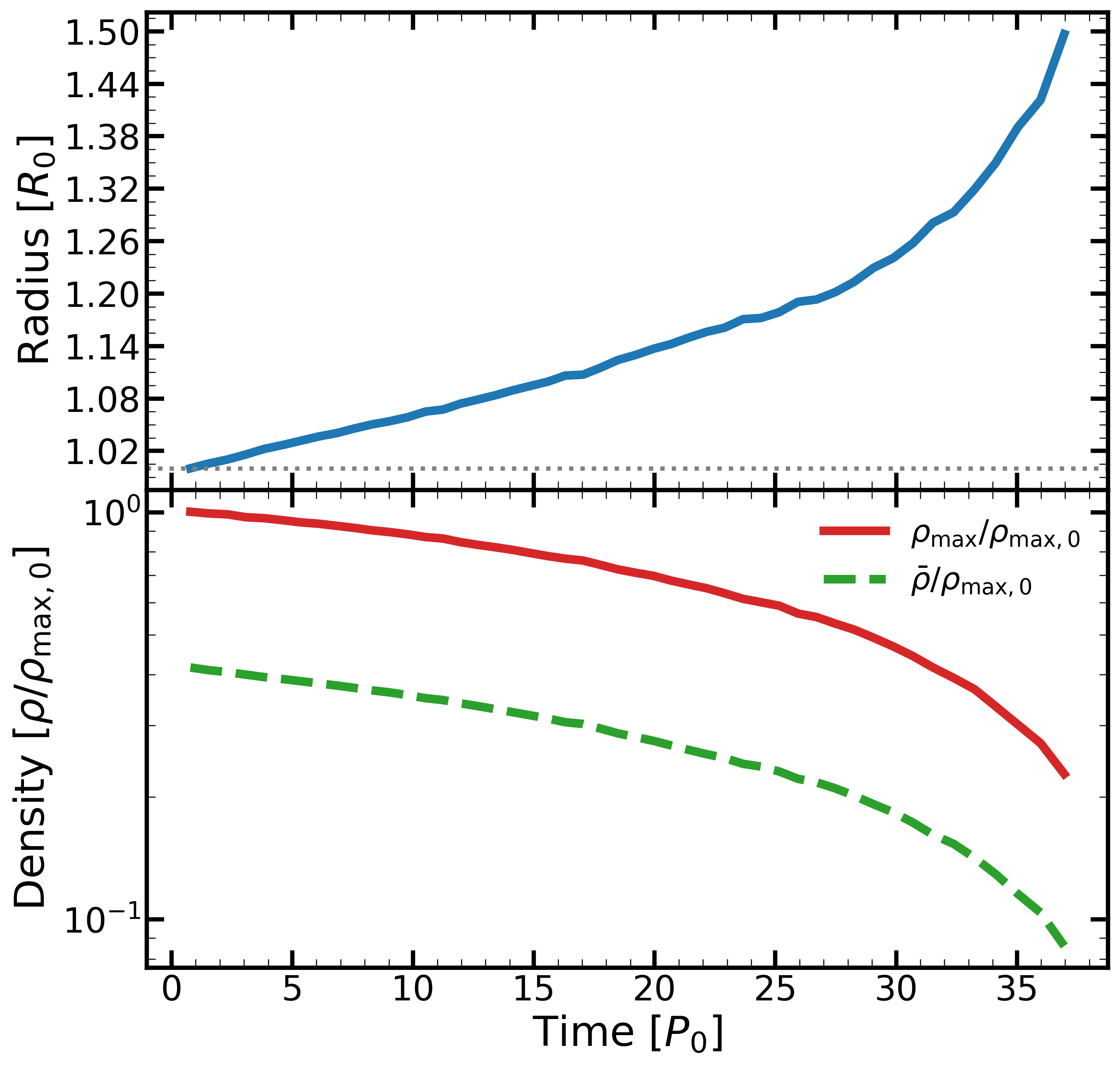}
		\caption{Structural response of the star during the tidal peeling process for Run A (see Fig.~\ref{fig:snapshots}). \textbf{Top Panel:} Time evolution of the stellar radius, $R_*\simeq R_{90}$, in units of $R_0$ (the initial stellar radius), defined as the radius enclosing 90\% of the bound mass. \textbf{Bottom Panel:} Time evolution of the mean and maximum stellar densities, in units of the initial maximum stellar density, $\rho_{\rm max,0}$. Both the mean density and core density decrease in time, consistent with adiabatic expansion.}
		\label{fig:structure_evolution}
	\end{figure}

	We explicitly demonstrate this behavior in Figure~\ref{fig:structure_evolution}, where we track the effective stellar radius $R_*(t)\simeq R_{90}$, defined as the Lagrangian radius enclosing 90\% of the bound mass (see Section~\ref{Analysis}). As mass stripping commences, $R_*(t)$ begins to increase monotonically, and both the mean density ($\bar{\rho}$) and the maximum core density ($\rho_{\max}$) decrease. This suggests that the stellar expansion is not limited to the outer envelope; rather, the entire star is adjusting to a new hydrostatic equilibrium with a larger radius. This result is consistent with the recent findings of \citet{2025ApJ...987...16B}, who demonstrated that low-mass stars---which are well-described by $\gamma=5/3$ polytropes---undergo a decrease in average density upon adiabatic mass loss, rendering them increasingly susceptible to further stripping.
	
\begin{figure}[ht!]
		\centering
		\includegraphics[width=1\linewidth]{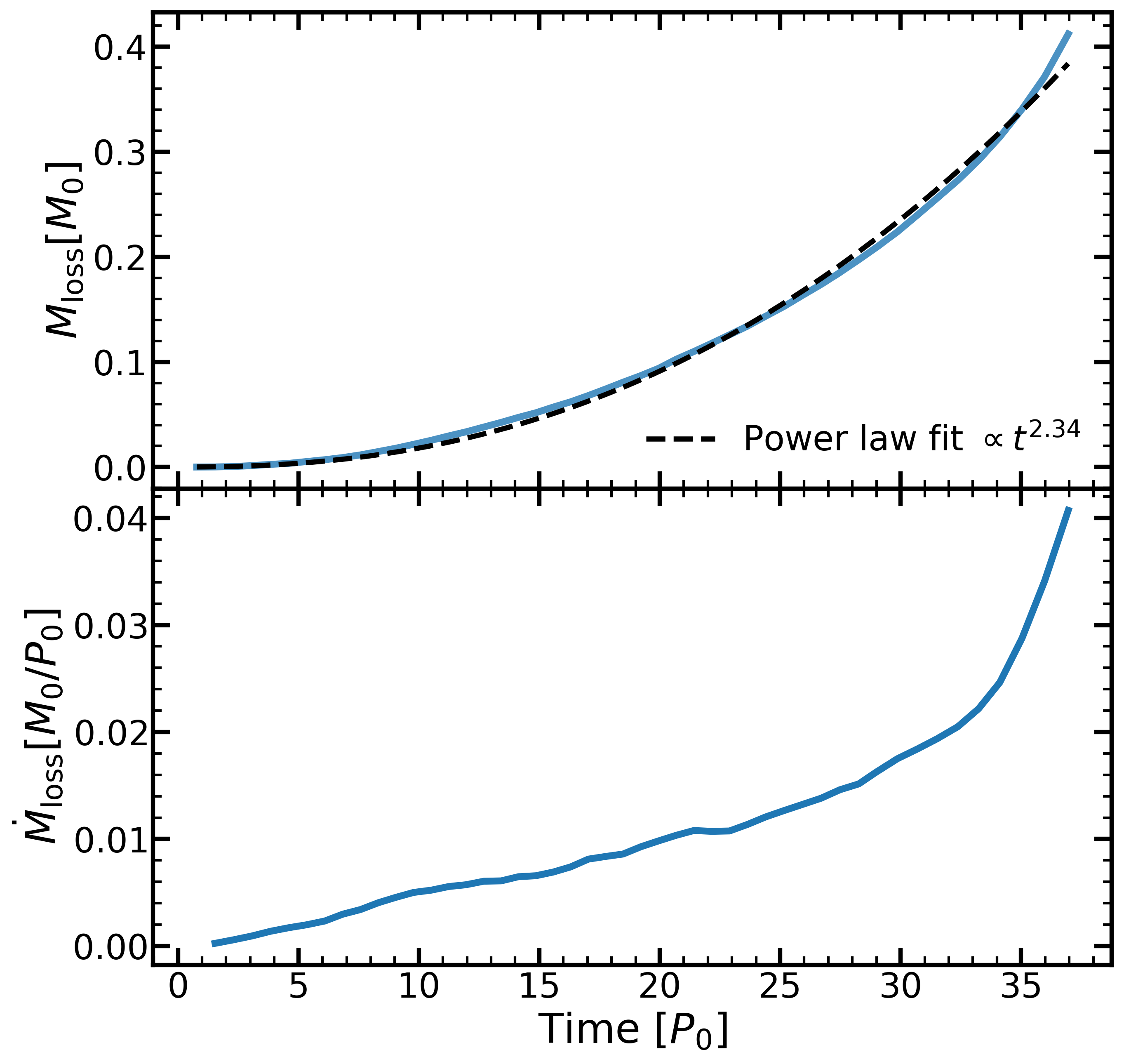}
		\caption{Evolution of stellar mass loss during tidal peeling for Run A (see Fig.~\ref{fig:snapshots}). \textbf{Top Panel:} Cumulative mass loss as a function of time. The dashed line indicates a power-law fit $M_{\rm loss}\propto t^{\alpha}$, with $\alpha=2.34$. \textbf{Bottom Panel:} Mass loss rate, $\dot{M}_{\rm loss}$. The acceleration of mass stripping is driven by the adiabatic expansion of the stellar envelope.}
		\label{fig:mass_history}
	\end{figure}

	The cumulative effect of orbital decay and stellar expansion leads to an increase in mass stripping. To quantify this process, we track both the cumulative mass loss from the star, $M_{\rm loss}(t)=M_0-M_*(t)$, and the instantaneous mass loss rate ($\dot{M}_{\rm loss}$), as presented in Figure~\ref{fig:mass_history}.
	The top panel shows that mass loss commences at the beginning, with the curve rising smoothly and accelerating. The cumulative mass loss during the runaway phase is well-described by a power-law:
	\begin{equation}
		M_{\rm loss}(t) \propto t^{\alpha} \quad (\alpha>1).
	\end{equation}
	This superlinear growth with $\alpha > 1$ implies an accelerating mass loss rate scaling as $\dot{M}_{\rm loss} \propto t^{\alpha-1}$, confirming the non-steady nature of the interaction. 
	
	\subsection{Orbital Decay and Circularization}
	
\begin{figure}[ht!]
		\centering
		\includegraphics[width=1\linewidth]{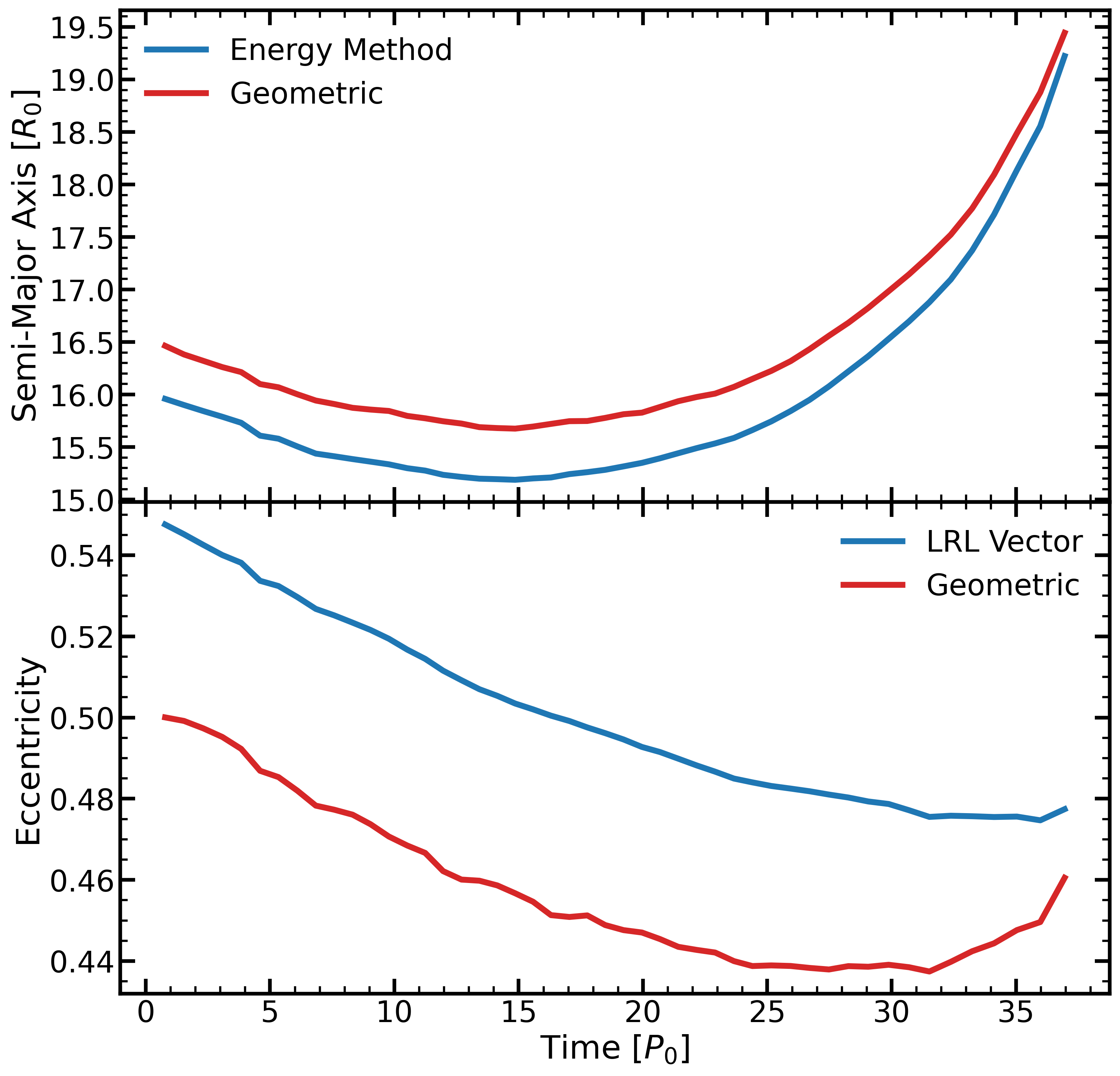}
		\caption{Evolution of the binary semi-major axis (SMA) and eccentricity for Run A (see Fig.~\ref{fig:snapshots}). In both panels, the blue solid lines denote the results obtained from the energy method (SMA) or Runge-Lenz vector method (eccentricity), while the red lines represent the results from the geometric method (see Section~\ref{Analysis}). Note the non-monotonic evolution of the SMA: the orbit initially decays due to tides, but turns around and expands at $t \approx 15 P_0$, driven by stellar mass loss.}
		\label{fig:orbitA}
	\end{figure}

	The tidal interaction extracts orbital energy and angular momentum from the binary motion, leading to an initial phase of orbital decay. However, the subsequent mass loss/transfer alters this trajectory, resulting in a non-monotonic orbital evolution driven by the competition between tidal dissipation and mass loss. To quantify this process, we plot the semi-major axis (SMA) and eccentricity as a function of time in Figure~\ref{fig:orbitA}. As discussed in Section~\ref{Analysis}, we use two different methods to determine the orbital SMA and eccentricity. The SMA initially shrinks due to tidal dissipation, reaching a minimum around $t \approx 15 P_0$. Then the orbit expands, coinciding with the mass loss phase. Concurrently, the eccentricity generally decreases but shows signs of flattening or a slight increase as the orbit expands. 

	For a binary with mass ratio $q = M_*/M_{\rm BH} = 0.1$, under the assumption of angular momentum conservation (or even non-conservative loss where the specific angular momentum of the ejecta is low), the loss of mass from the lighter component necessitates an expansion of the orbit. Overall, the rate of change of the SMA can be written as the sum of two terms:
	\begin{equation}
		\frac{\dot{a}}{a} = \left( \frac{\dot{a}}{a} \right)_{\rm tides} + \left( \frac{\dot{a}}{a} \right)_{\rm mass\,loss}
	\end{equation}
	where the first term is negative and the second term is positive. The turnover point observed in our simulation corresponds to the moment where the mass stripping effect overwhelms the tidal decay rate.
	
	\subsection{Runaway Instability Mechanism} \label{subsec:instability_mech}
	
	The catastrophic disruption of the star is driven by the breakdown of mass loss/transfer stability. To quantify this instability, we must compare the structural response of the star with the evolution of its Roche lobe. We define the adiabatic mass-radius exponent of the star, $\zeta_{\rm ad}$, and the response of the Roche lobe, $\zeta_L$, by:
	\begin{equation}
		\zeta_{\rm ad} \equiv  \frac{d \ln R_*}{d \ln M_*} , \quad \zeta_L \equiv \frac{d \ln R_L}{d \ln M_*}.
		\label{eq:zeta}
	\end{equation}
	Mass transfer is stable only if the star contracts relative to the Roche lobe, i.e., $\zeta_{\rm ad} \ge \zeta_L$. Conversely, if $\zeta_{\rm ad} < \zeta_L$, the star expands faster than its Roche lobe, leading to dynamical instability.
	
\begin{figure}[ht!]
		\centering
		\includegraphics[width=1\linewidth]{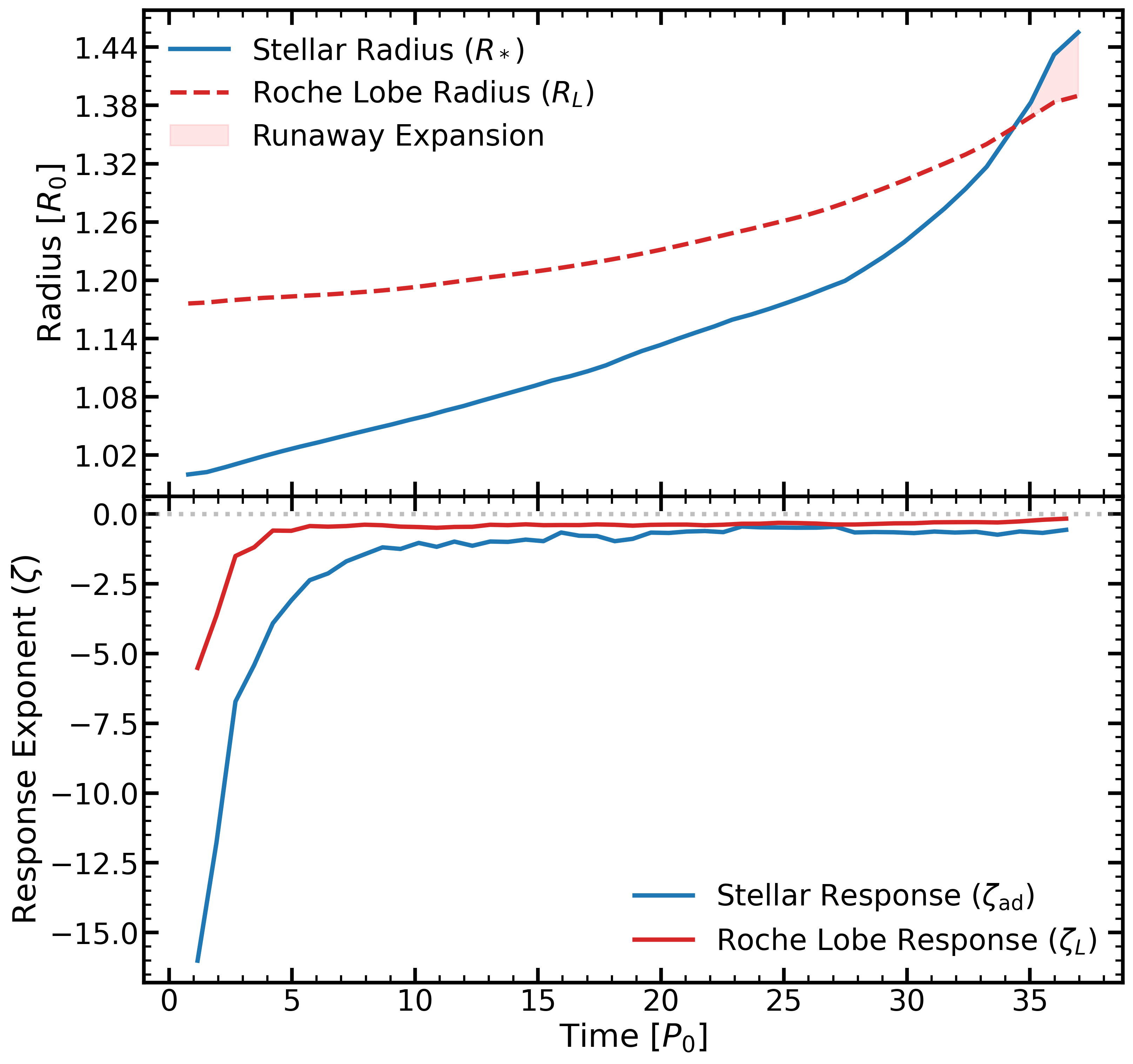}
		\caption{Dynamical mechanism of runaway mass stripping for Run A. \textbf{Top Panel:} Evolution of the stellar radius, $R_*\simeq R_{90}$, in units of $R_0$ (the initial stellar radius), and the Roche lobe radius (see Eq.~\ref{eq:roche}). The stellar radius becomes larger than $R_L$ at $t \approx 34 P_0$, marking the onset of the runaway phase (shaded pink area). \textbf{Bottom Panel:} Evolution of the response exponents (see Eq.~\ref{eq:zeta}). The stellar adiabatic response ($\zeta_{\rm ad}$, blue) is consistently more negative than the Roche lobe response ($\zeta_L$, red).}
		\label{fig:zeta_evolution}
	\end{figure}

	Figure~\ref{fig:zeta_evolution} shows the evolution of $R_*$, $R_L$, $\zeta_{\rm ad}$, and $\zeta_L$ for Run A. Initially ($t \lesssim 30 P_0$), the star resides within its Roche lobe ($R_* < R_L$), although both radii increase over time: $R_L$ due to pericenter expansion and $R_*$ due to mass loss. Since $\zeta_{\rm ad} < \zeta_L$ holds, the stellar radius grows at a significantly steeper rate than the Roche lobe. This differential growth leads to the crossover of $R_*$ and $R_L$ at $t \simeq 34 P_0$, where the star overtakes its Roche lobe ($R_* > R_L$). Beyond this point, the system enters the runaway phase, leading to the eventual disruption of the star.

\begin{figure}[ht!]
		\centering
		\includegraphics[width=1\linewidth]{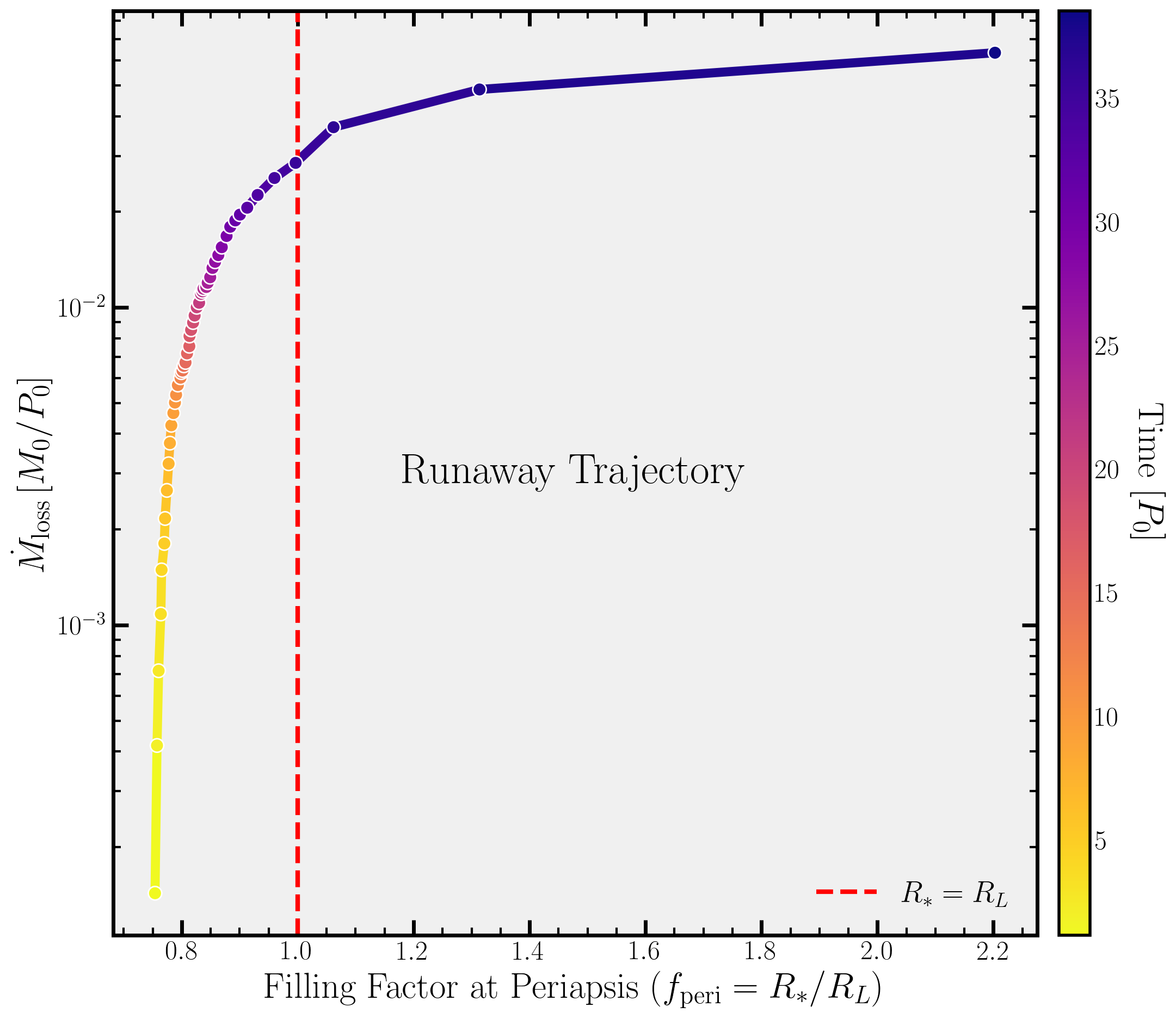}
		\caption{Phase space trajectory of the runaway mass-loss instability for Run A. We plot the mass loss rate $\dot{M}_{\rm loss}$ against the Roche lobe filling factor at pericenter ($f_{\rm peri} \equiv R_* / R_L$). The points are colored by time. The red dashed line marks $f_{\rm peri}=1$. A stable system would find an equilibrium near $f_{\rm peri} \approx 1$ with a constant $\dot{M}_{\rm loss}$. Instead, our simulation shows a clear \textit{runaway trajectory} moving upward and to the right: mass loss triggers adiabatic stellar expansion, which increases the filling factor, which in turn drives higher mass loss rates.}
		\label{fig:phase_diagram}
	\end{figure}

	To distinguish the runaway disruption from Roche-lobe overflow (RLOF), we further analyze the system's trajectory in the phase space of mass loss rate ($\dot{M}_{\rm loss}$) versus the filling factor $f_{\rm peri} = R_*/R_L$, evaluated at pericenter (Fig.~\ref{fig:phase_diagram}). In a self-regulating, stable RLOF scenario, the system is expected to find an equilibrium where $f_{\rm peri} \approx 1$ and maintain a steady mass transfer rate.
	
	In contrast, our simulation reveals an unstable behavior. Initially, $f_{\rm peri} < 1$ and the star remains detached until orbital decay drives $f_{\rm peri}$ toward unity, initiating mass loss. Once mass transfer commences and $f_{\rm peri}$ exceeds unity, a positive feedback loop is triggered. The mass loss causes the convective envelope to expand ($\zeta_{\rm ad} < 0$), while the Roche lobe fails to expand sufficiently to accommodate it. This forces $f_{\rm peri}$ to increase continuously despite the mass loss, which in turn drives higher mass loss rate and faster stellar expansion.
	
	\subsection{Disk Formation} \label{subsec:disk_structure}
	
\begin{figure}[ht!]
		\centering
		\includegraphics[width=1\linewidth]{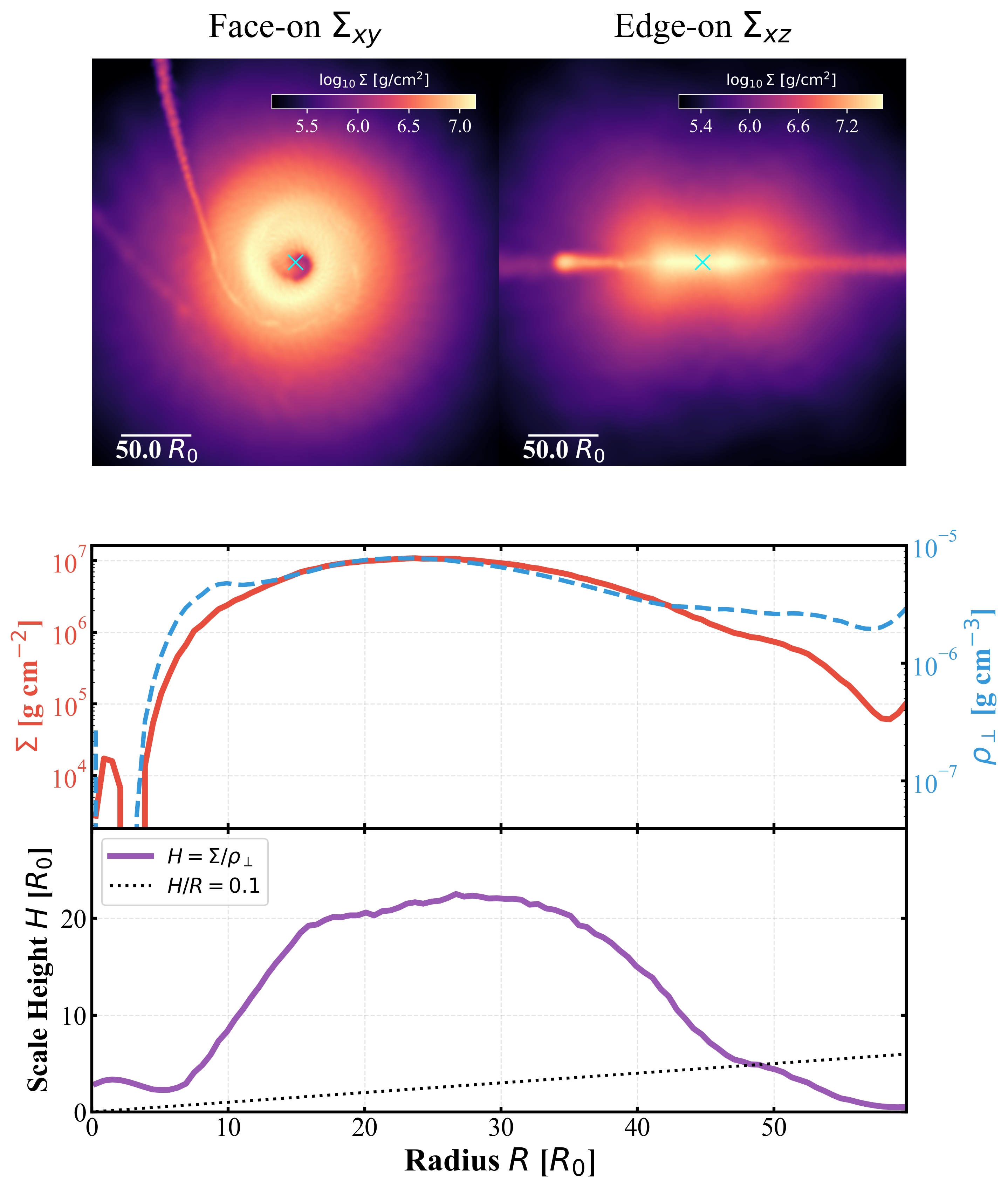}
		\caption{Geometric and structural analysis of the accretion torus formed after stellar disruption at the end of the simulation ($t = 60 P_0$) for Run A. 
			\textbf{Top Panels:} Face-on ($\Sigma_{xy}$) and edge-on ($\Sigma_{xz}$) surface density maps, showing a coherent toroidal structure after circularization. 
			\textbf{Middle Panel:} Azimuthally-averaged radial profiles of the surface density $\Sigma$ (solid red, left axis) and the midplane volume density $\rho_\perp$ (dashed blue, right axis). 
			\textbf{Bottom Panel:} Radial profile of the scale height $H = \Sigma / \rho_\perp$ (solid purple line). The dotted line represents constant aspect ratios of $H/R = 0.1$. The scale height significantly exceeds $H/R = 0.7$ in the main body of the disk.}
		\label{fig:disk_geometry}
	\end{figure}

	Following the runaway mass loss, the star is effectively shredded by the tidal force. As shown in the top panels of Fig.~\ref{fig:disk_geometry}, the stellar debris intersects with itself, dissipating orbital energy through strong shocks and settling into a toroidal configuration around the BH.
	
	To quantify this flow, we isolate the bound, circularized fluid elements from the extended tidal tails by applying a kinematic cut, retaining only fluid particles with an instantaneous eccentricity $e \lesssim 0.45$. Fig.~\ref{fig:disk_geometry} shows the disk structure at the terminal stage of our simulation ($t \simeq 60 P_0$).
	
	We define the azimuthally-averaged surface density $\Sigma(R)$ as:
	\begin{equation}
		\Sigma(R) = \frac{1}{2\pi} \int_0^{2\pi} \int_{-\infty}^{\infty} \rho(R, \phi, z) dz d\phi.
	\end{equation}
	and the midplane density $\rho_\perp(R)$ as:
	\begin{equation}
		\rho_\perp(R) = \frac{1}{2\pi} \int_0^{2\pi} \rho(R, \phi, z=0) d\phi.
	\end{equation}
	The scale height of the accretion flow is then:
	\begin{equation}
		H(R) = \frac{\Sigma(R)}{\rho_\perp(R)}.
	\end{equation}
	
	Fig.~\ref{fig:disk_geometry} reveals that the debris has settled into a geometrically thick torus. Both $\Sigma(R)$ and $\rho_\perp(R)$ peak at $R = 25-30\,R_0$ (recall $R_0$ is the initial radius of the star). The scale height $H(R)$ increases with radius and reaches a maximum of $\sim 24\,R_0$ near the density peak. Throughout the bulk of the disk ($R \lesssim 45\,R_0$), the aspect ratio $H/R$ remains above 0.1.
	
	This geometric thickening is a direct consequence of shock heating that precedes disk formation. In our simulation, the heat generated by stream-stream intersection shocks and the fallback of high-entropy material is trapped within the flow. The resulting thermal pressure counterbalances the vertical component of gravity, inflating the debris into a pressure-supported torus. Such an optically thick configuration may give rise to soft X-ray to UV emission associated with such micro-TDEs.
\subsection{Black-Hole Accretion} \label{sec:lightcurve}

To characterize the potential observable signature of the runaway stellar disruption, we analyze the mass supply/accretion rate $\dot{M}_{\rm acc}$ onto the BH. We calculate $\dot{M}_{\rm acc}$ by tracking the flux of SPH particles crossing the sink radius ($r_{\rm sink}=500\,GM_{\rm BH}/c^2$) at every timestep. This rate is normalized by the Eddington accretion rate, defined as:
\begin{equation}
	\dot{M}_{\rm Edd} \equiv \frac{L_{\rm Edd}}{\eta c^2} = \frac{4\pi G M_{\rm BH} m_p c}{\eta \sigma_T c^2} \, ,
\end{equation}
where $L_{\rm Edd} = 1.26 \times 10^{39} (M_{\rm BH}/10\,M_\odot) \, \rm erg \, s^{-1}$ is the Eddington luminosity, $m_p$ is the proton mass, $\sigma_T$ is the Thomson cross-section, and we assume a standard radiative efficiency of $\eta = 0.1$.

	\begin{figure}[ht!]
		\centering
		\includegraphics[width=1\columnwidth]{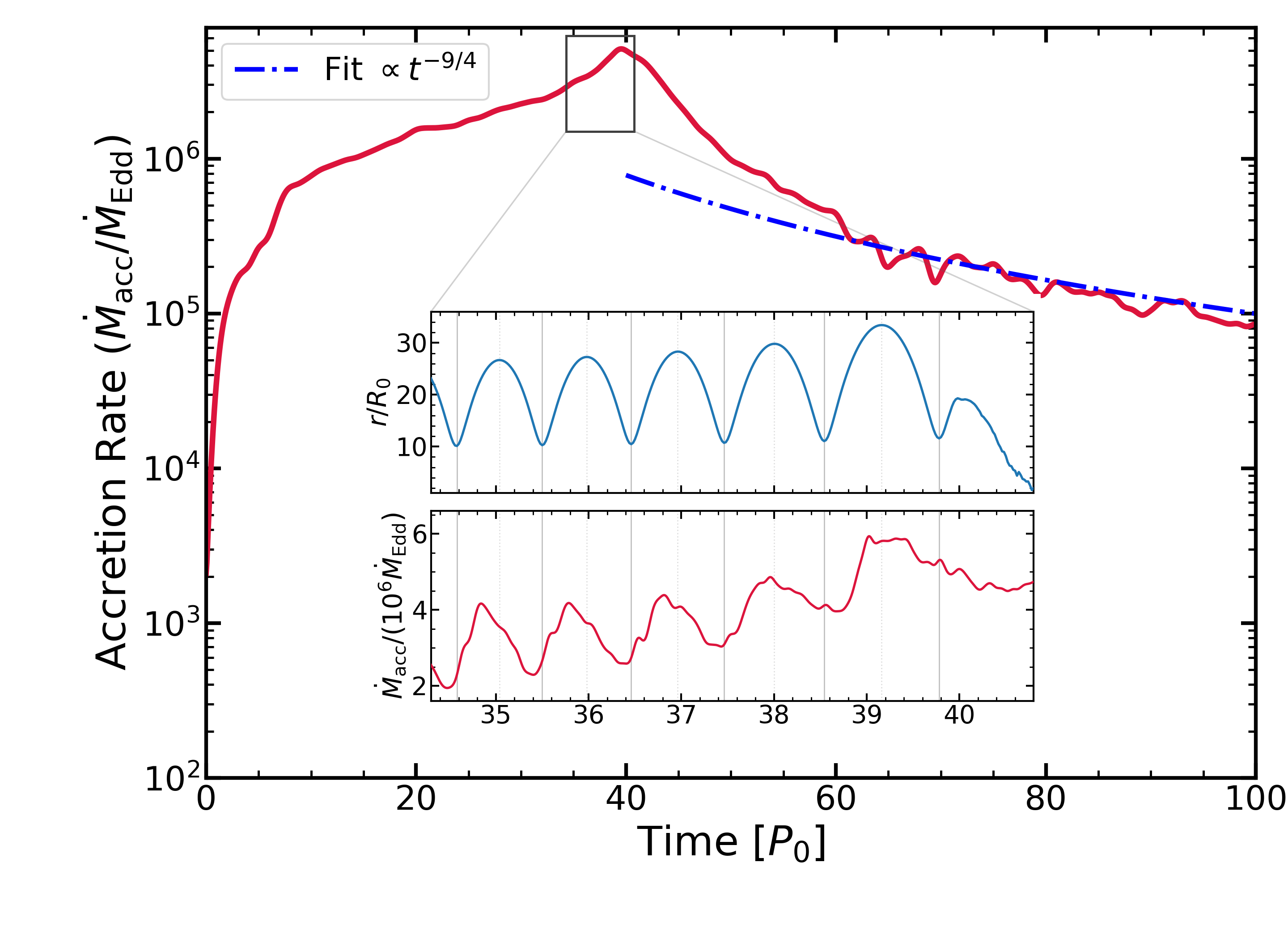}
	\caption{
		Evolution of the mass supply/accretion rate $\dot{M}_{\rm acc}$ onto the BH for Run A. The dimensionless accretion rate ($\dot{m}_{\rm acc} \equiv \dot{M}_{\rm acc}/\dot{M}_{\rm Edd}$) is derived from the cumulative mass crossing the sink radius around the BH. The late-time decay is approximately described by $\dot{m}_{\rm acc} \propto t^{-9/4}$. The inset zooms in on the final pre-disruption cycles. In the inset, $r$ denotes the binary separation.
	}
	\label{fig:lightcurve}
\end{figure}

Fig.~\ref{fig:lightcurve} shows the BH accretion history in our simulation. Prior to stellar disruption, the accretion rate exhibits quasi-periodic modulation, with the peaks and valleys shifted from the pericenter passages. The accretion rate reaches a maximum ($\dot{m}_{\rm acc}\sim5\times 10^6$) at $t\sim 40 P_0$, corresponding to the onset of stellar disruption. Following the maximum, the accretion rate decays according to a power-law $\dot{m}_{\rm acc} \propto t^{-9/4}$. This steep decay index resembles that of partial TDEs involving a surviving stellar core \citep{2019ApJ...883L..17C,2020ApJ...899...36M}, and is distinct from the canonical $t^{-5/3}$ fallback rate expected for full TDEs involving supermassive BHs. In our case, the stellar disruption occurs at modest eccentricity ($e \sim 0.48$; see Fig.~\ref{fig:orbitA}) and has a smaller $M_{\rm BH}/M_*$ than TDEs involving supermassive BHs, so the $t^{-5/3}$ scaling does not apply. The extreme super-Eddington rate ($\dot{m}_{\rm acc}\gg 1$) shown in Fig.~\ref{fig:lightcurve} is consistent with the disk structure depicted in Fig.~\ref{fig:disk_geometry}.
	At such a high accretion rate, the trapped radiation pressure inflates the flow vertically, creating a geometrically thick structure. Overall, the accretion history shown in Fig.~\ref{fig:lightcurve} provides a consistent picture of a super-Eddington, pressure-supported flow produced by the rapid stripping of the stellar envelope.

	\subsection{Numerical Robustness}\label{subsec:numerical_robustness}

\begin{figure}[htbp!]
		\centering
		\includegraphics[width=1\columnwidth]{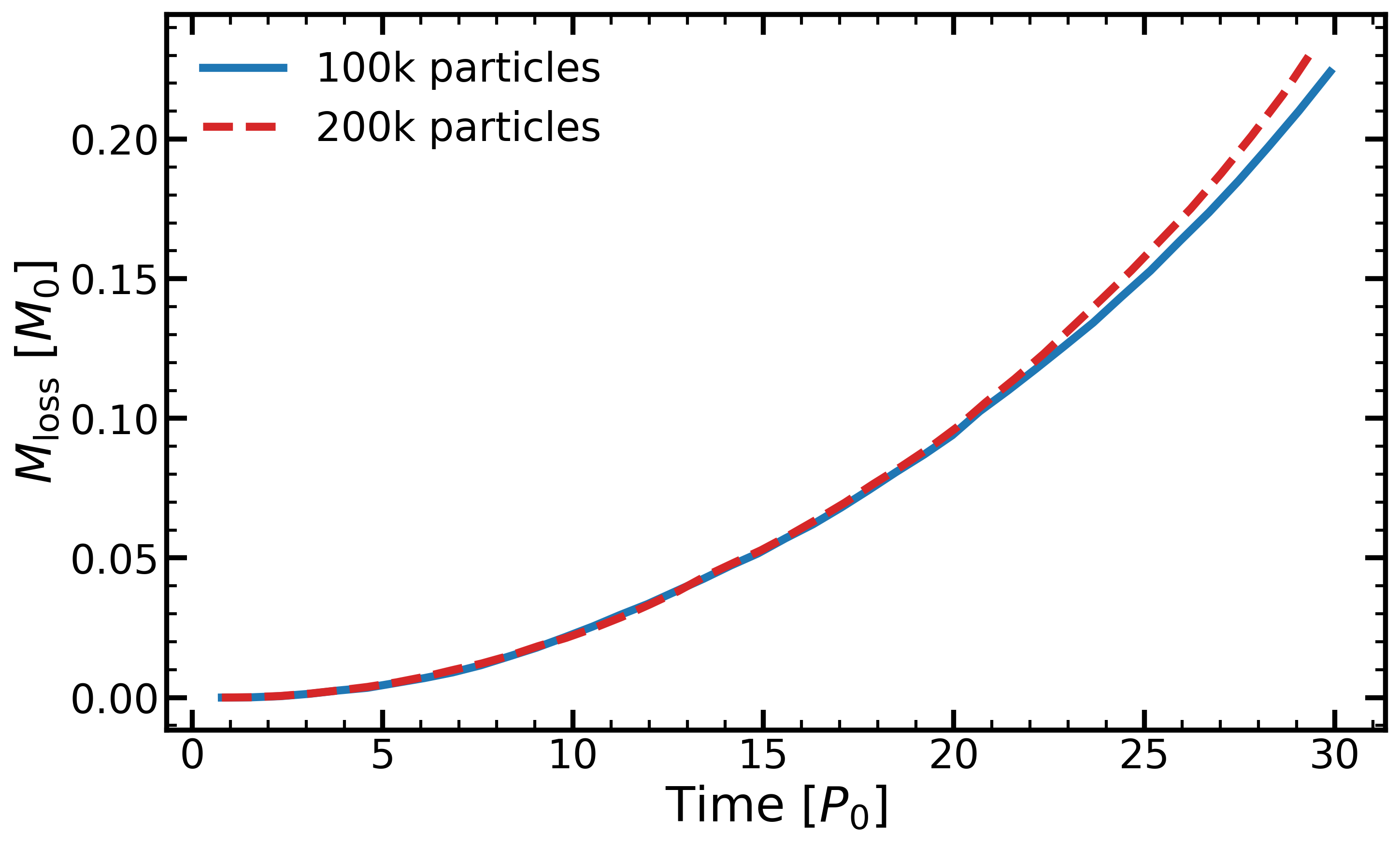}
		\caption{Robustness test comparing the mass loss evolution in simulations with $10^5$ SPH particles Run A and $2\times 10^5$ particles.}
		\label{fig:robustness}
	\end{figure}

	Our fiducial Run A and the other runs presented in this paper use $N=10^5$ SPH particles. To test the numerical robustness of our results, we repeated Run A using $N=2 \times 10^5$ particles. Fig.~\ref{fig:robustness} shows that the mass loss evolution is nearly identical at both resolutions. This agreement suggests that our fiducial SPH resolution ($N=10^5$) is sufficient to accurately capture the mass loss/transfer process and the subsequent runaway instability.

	\section{Results and Analysis for Run B: \texorpdfstring{\lowercase{$b_0=3.57$, $e_0=0.55$}}{b0=3.57, e0=0.55}} \label{sec:results2}

\begin{figure}[ht!]
	\centering
	\includegraphics[width=1\columnwidth]{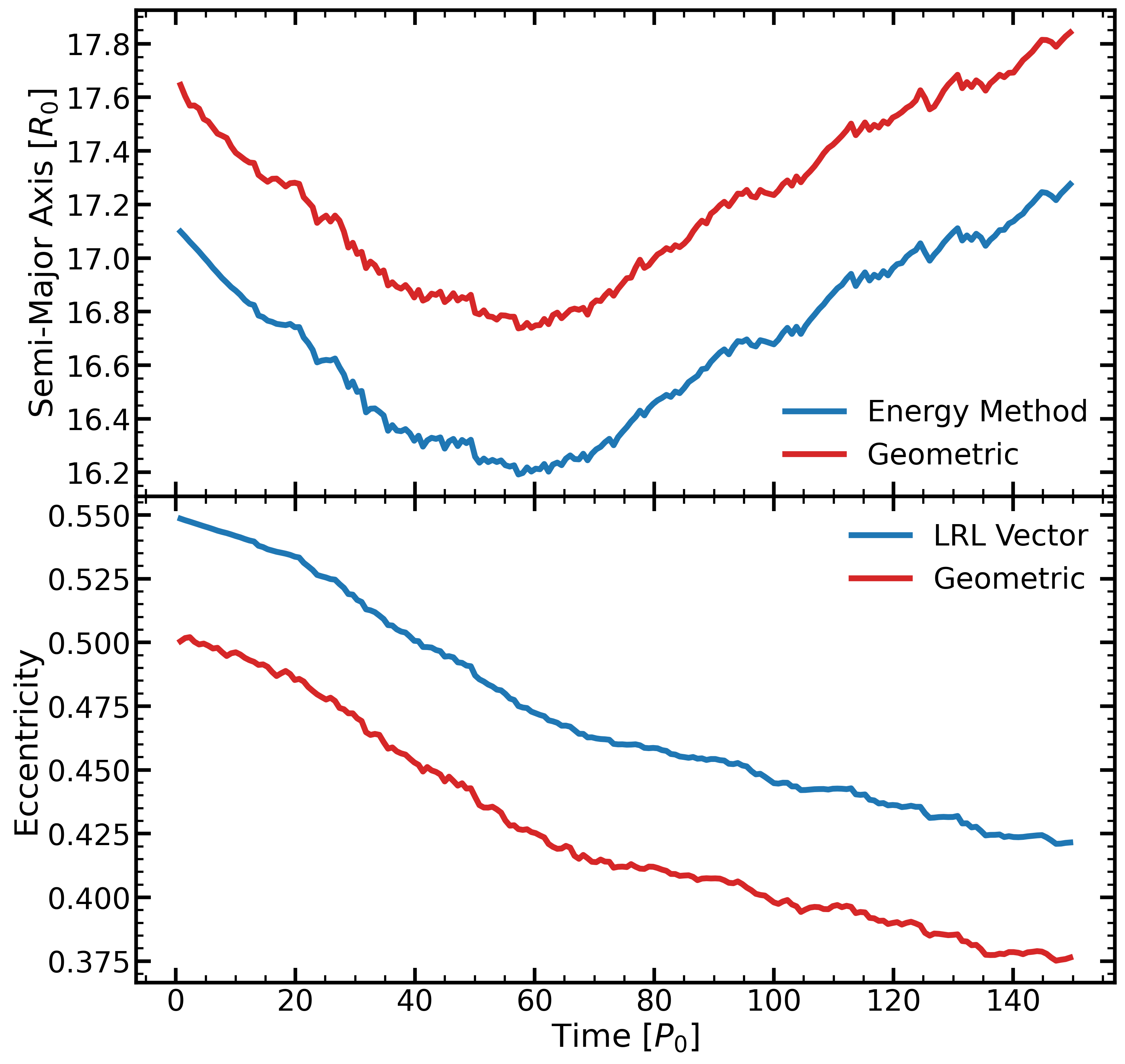}
	\caption{Same as Fig.~\ref{fig:orbitA}, except for Run B (with initial $b_0=3.57$).}
	\label{fig:caseB_orbit}
\end{figure}

\begin{figure}[ht!]
	\centering
	\includegraphics[width=1\columnwidth]{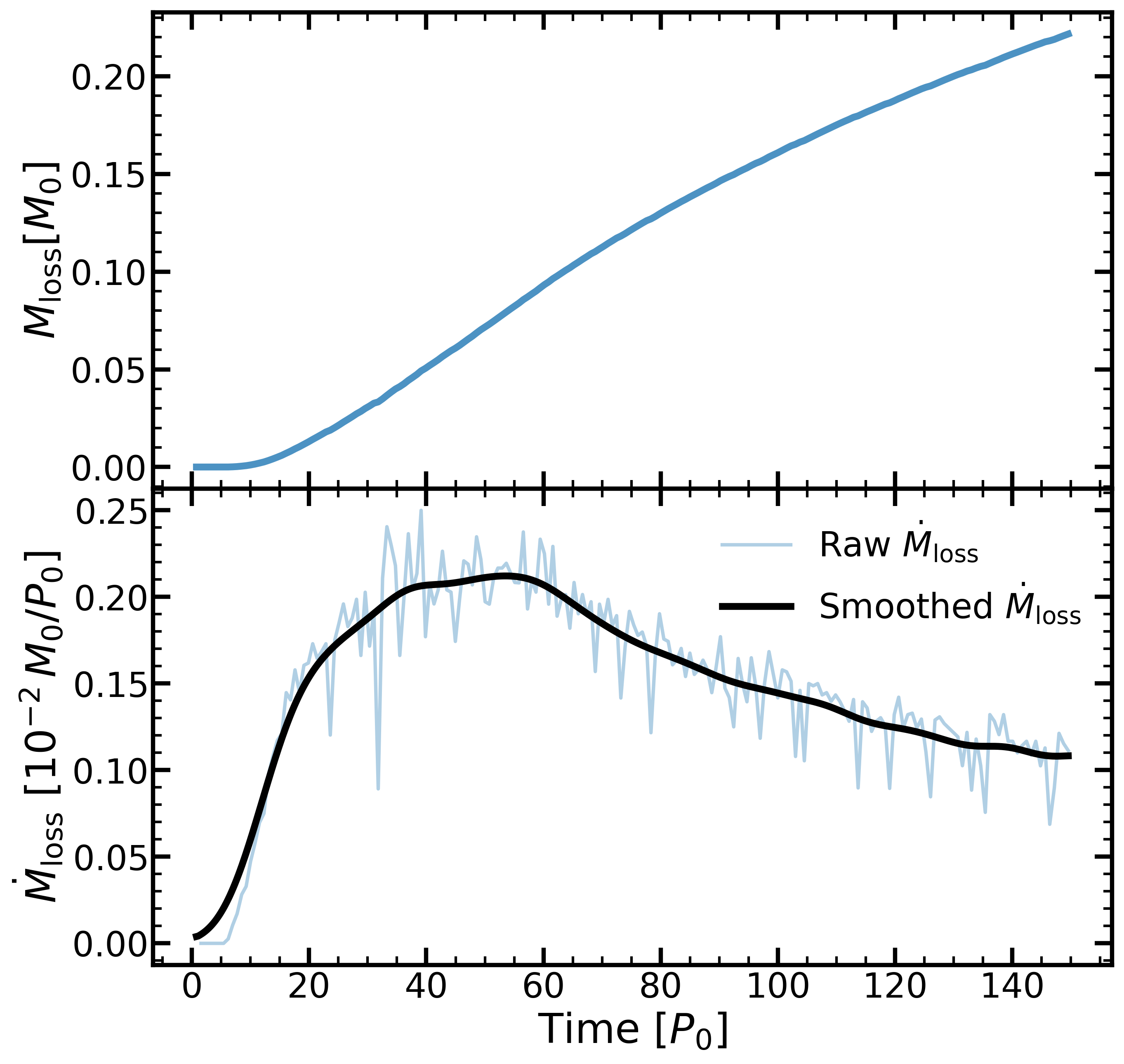}
	\caption{Same as Fig.~\ref{fig:mass_history}, except for Run B. The cumulative mass loss reaches $20\%$ over $150 P_0$, showing a linear or sublinear growth at late times rather than the exponential runaway seen in Run A. In the bottom panel, the black curve represents the smoothed time derivative of the cumulative mass loss.}
	\label{fig:caseB_mass}
\end{figure}

While Run A demonstrates the runaway stellar disruption, Run B (with initial $b_0=3.57$, or $\beta_0=0.28$) illustrates the case of stable mass transfer on an eccentric orbit, revealing the sensitivity of the system to the initial impact parameter. Fig.~\ref{fig:caseB_orbit} shows the evolution of the binary semi-major axis and eccentricity over 150 $P_0$. Fig.~\ref{fig:caseB_mass} shows the corresponding stellar mass loss history. The evolution exhibits two distinct phases:

\textbf{Phase I: Tidal Decay ($0 < t \lesssim 60 P_0$).} Initially, the system undergoes a prolonged phase of orbital decay, and the semi-major axis shrinks from $17.1 R_0$ to about $16.2 R_0$, driven purely by the dissipation of dynamical tides. The eccentricity decays monotonically from $0.55$ to $\approx 0.42$, indicating significant circularization. Stellar mass loss is negligible for $t \lesssim 7 P_0$, and the mass loss rate gradually increases for $t \gtrsim 10 P_0$, reaching a peak around $t\sim 50 P_0$. 

\textbf{Phase II: Mass-Loss-Induced Expansion ($t \gtrsim 60 P_0$).} As mass loss intensifies, the orbital semi-major axis reverses its decline and begins to expand. This turnaround occurs significantly later than in Run A, suggesting that mass loss must be sufficiently large to be dynamically relevant.

Importantly, instead of accelerating toward disruption, the mass loss rate rolls over and begins to decline after $t \approx 50 P_0$. The cumulative mass loss transitions from an exponential-like onset to a linear or sublinear growth, resulting in $\sim 20\%$ of total mass loss by $t = 150 P_0$. This behavior suggests that the system has found a quasi-stable equilibrium where the orbital expansion (driven by mass loss) sufficiently counteracts the stellar expansion, preventing the catastrophic deepening of the contact.

\begin{figure}[htbp]
	\centering
	\includegraphics[width=1\columnwidth]{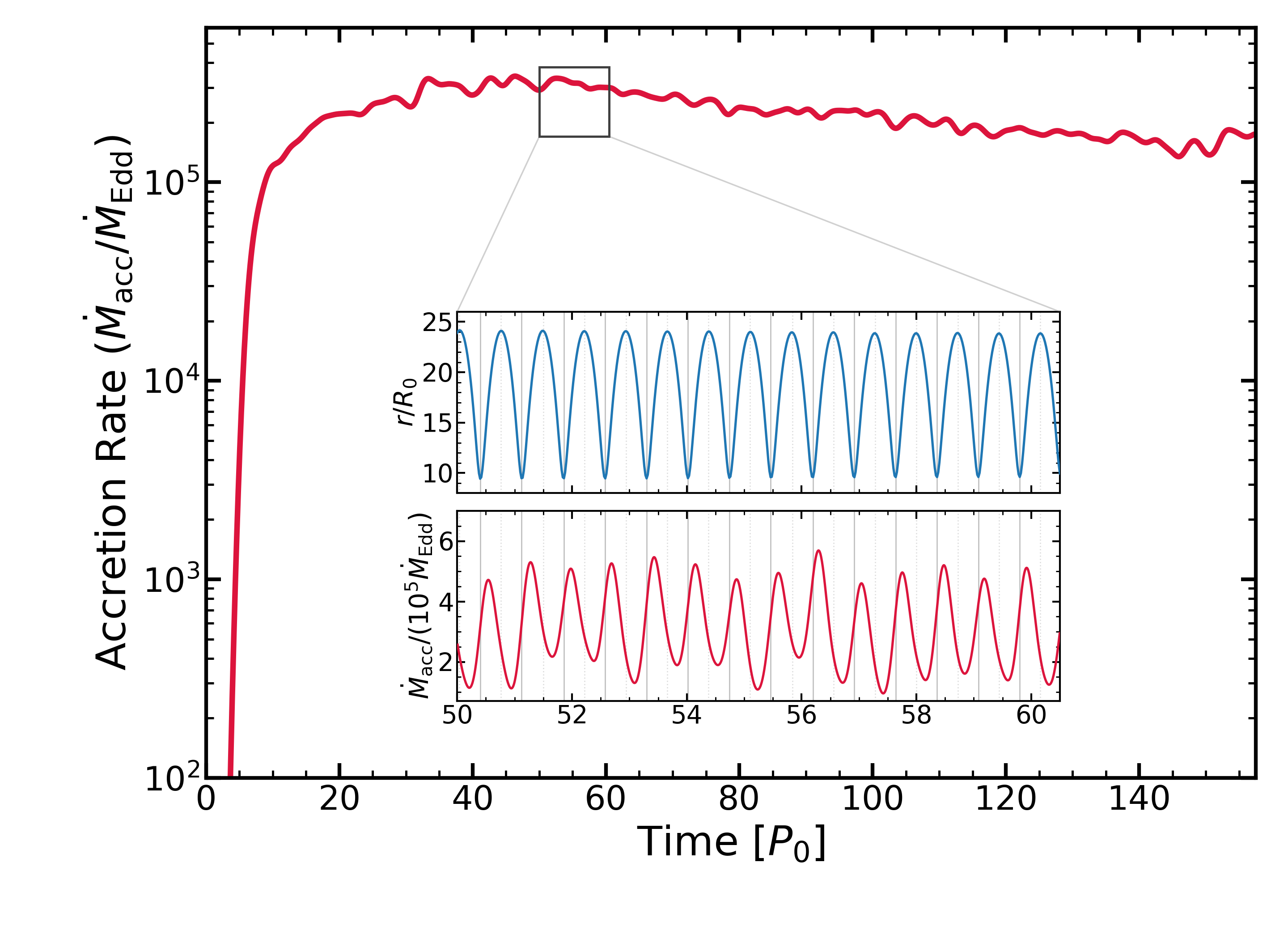}
	\caption{Same as Fig.~\ref{fig:lightcurve}, but for Run B. The inset highlights a representative interval of quasi-periodic modulation in the regulated mass-transfer phase.}
	\label{fig:caseB_lightcurve}
\end{figure}

The corresponding BH accretion history reinforces this interpretation. Fig.~\ref{fig:caseB_lightcurve} shows that, after the onset of mass transfer, the accretion rate rises to a few $\times 10^5\,\dot{M}_{\rm Edd}$, but remains below the peak reached in Run A (see Fig.~\ref{fig:lightcurve}). The curve forms a broad maximum and then declines slowly. Thus, Run B still supplies the BH efficiently, but the feeding is sustained by self-regulated mass transfer rather than by a terminal stellar disruption.

\begin{figure}[htbp]
	\centering
	\includegraphics[width=1\linewidth]{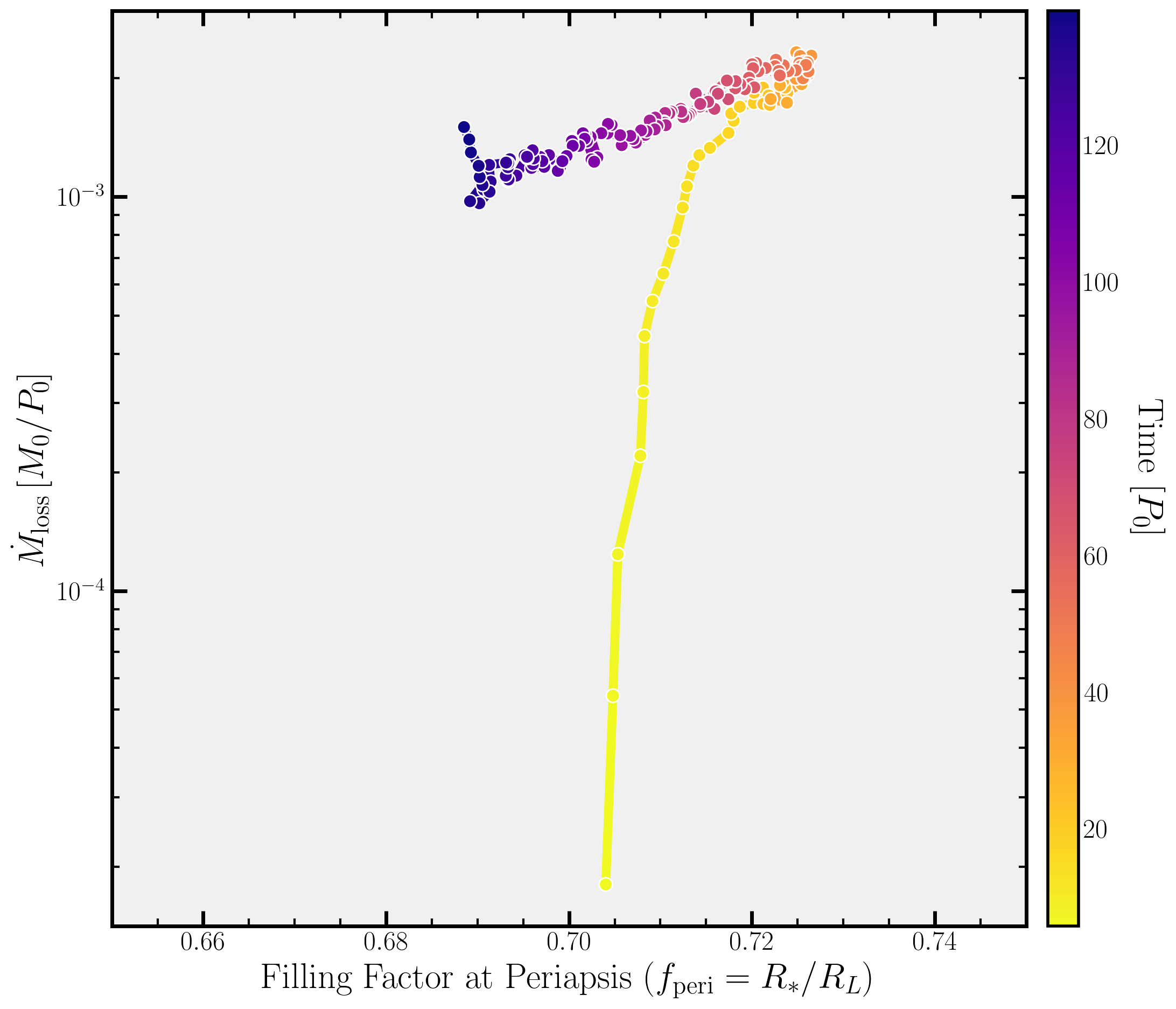}
	\caption{Same as Fig.~\ref{fig:phase_diagram}, except for Run B.}
	\label{fig:phase_space_B}
\end{figure}

Fig.~\ref{fig:phase_space_B} shows the mass loss trajectory in the $\dot{M}_{\rm loss} - f_{\rm peri}$ diagram. In contrast to Fig.~\ref{fig:phase_diagram} (for Run A), the topological features of Fig.~\ref{fig:phase_space_B} explicitly capture the self-regulating nature of Run B.

During the early mass transfer phase ($t \sim 20$--$40 P_0$), the trajectory moves upward and to the right. The onset of mass loss triggers the adiabatic expansion of the stellar envelope ($\zeta_{\rm ad} < 0$), causing the filling factor to increase and driving up the mass loss rate. However, unlike the monotonic runaway observed in Run A, this trajectory does not cross the critical $R_* = R_L$ boundary. Instead, the filling factor reaches a maximum of $f_{\rm peri}\approx0.73$ at $t\simeq 50 P_0$. At this point, the orbital widening driven by the mass loss (which acts to increase $R_L$) begins to outpace the adiabatic expansion of the star ($R_*$), and the trajectory moves to the left, with a decreasing filling factor. At late times ($t \gtrsim 100 P_0$), the system stagnates into a dense cluster around $f_{\rm peri} \approx 0.68$--$0.70$ with a steady mass loss rate. This closed-loop topology demonstrates that the system has found a quasi-stable equilibrium where negative feedback from orbital expansion successfully neutralizes the positive feedback of adiabatic stellar expansion, preventing the catastrophic stellar disruption.

\section{Other Simulation Runs: Threshold for Runaway Stellar Disruption}
\label{sec:threshold}

\setcounter{topnumber}{4}
\setcounter{bottomnumber}{4}
\setcounter{totalnumber}{6}
\renewcommand{\topfraction}{0.95}
\renewcommand{\bottomfraction}{0.95}
\renewcommand{\textfraction}{0.05}

The contrasting outcomes of Run A ($b_0=3.33$) and Run B ($b_0=3.57$) reveal two possible evolutionary pathways for eccentric star-BH binaries: runaway tidal disruption and long-lived, self-regulated mass transfer. The boundary between these outcomes depends on the combined effects of the initial pericenter distance and eccentricity. To explore this dependence, we carry out four additional simulations (Runs C--F; see Table~\ref{tab:initial_conditions}).

\begin{figure}[htbp]
	\centering
	\includegraphics[width=0.88\linewidth]{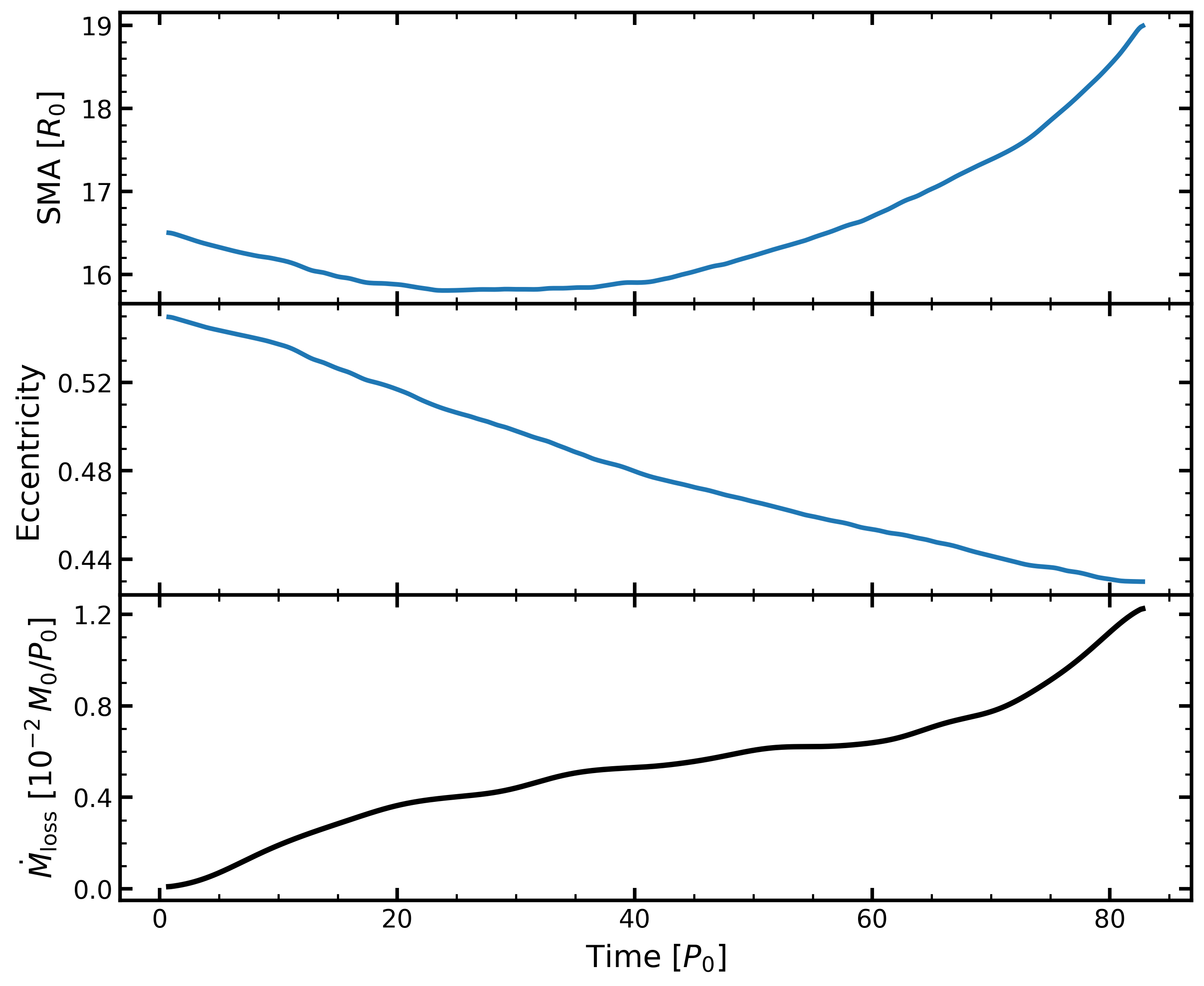}
	\caption{Results for Run C (with $b_0=3.45$ and $e_0=0.55$). The panels show the semi-major axis (from the energy method), eccentricity (from the Runge-Lenz vector method), and smoothed mass-loss rate evolution. The cumulative mass loss exceeds $0.4\,M_0$ at $t\simeq 85 P_0$, and the star is subsequently disrupted.}
	\label{fig:runC_summary}
\end{figure}

\begin{figure}[htbp]
	\centering
	\includegraphics[width=0.88\linewidth]{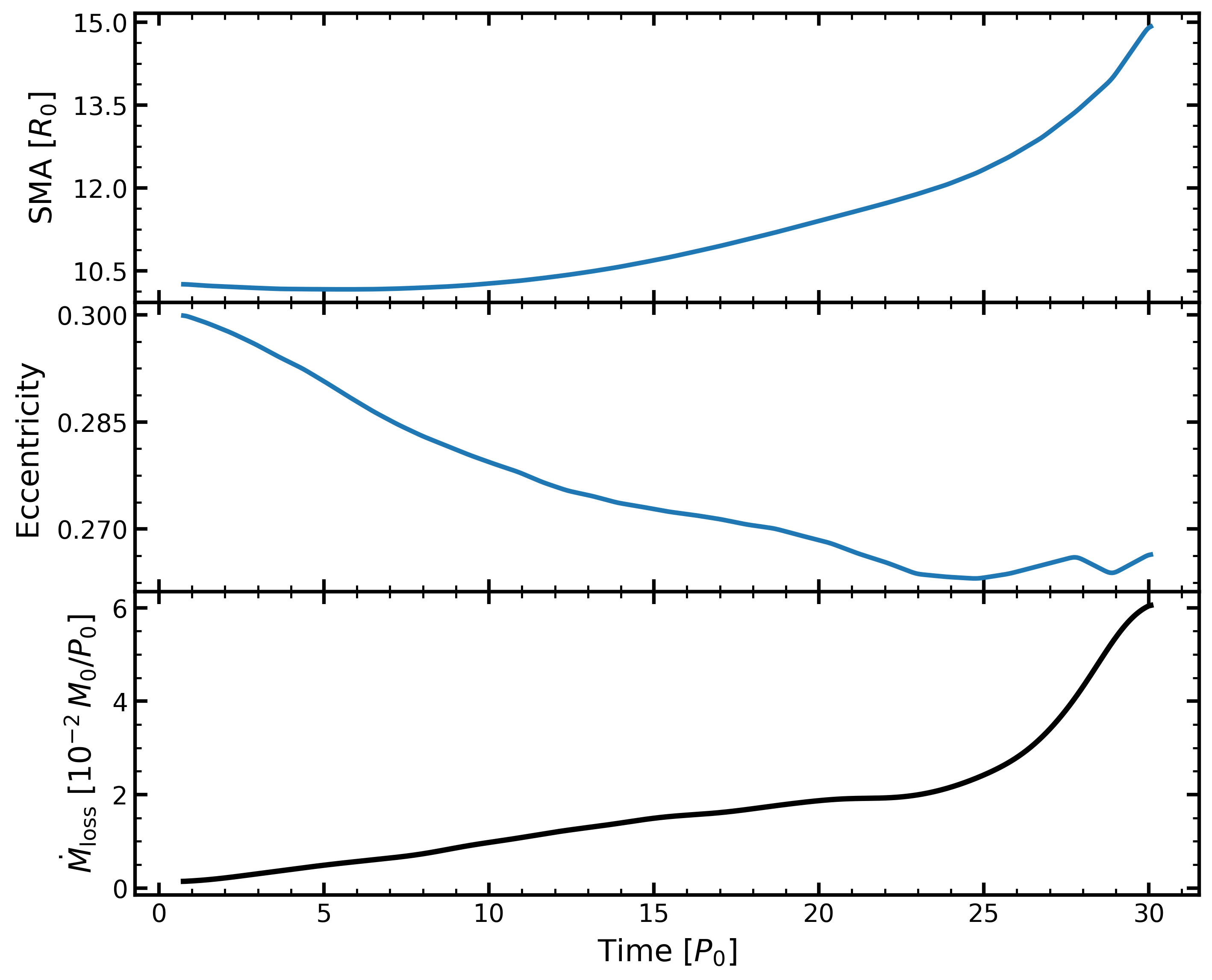}
	\caption{Results for Run D (with $b_0=3.33$ and $e_0=0.30$). The cumulative mass loss reaches about $0.6\,M_0$ by $t\simeq 30 P_0$, followed by disruption.}
	\label{fig:runD_summary}
\end{figure}

\begin{figure}[htbp]
	\centering
	\includegraphics[width=0.88\linewidth]{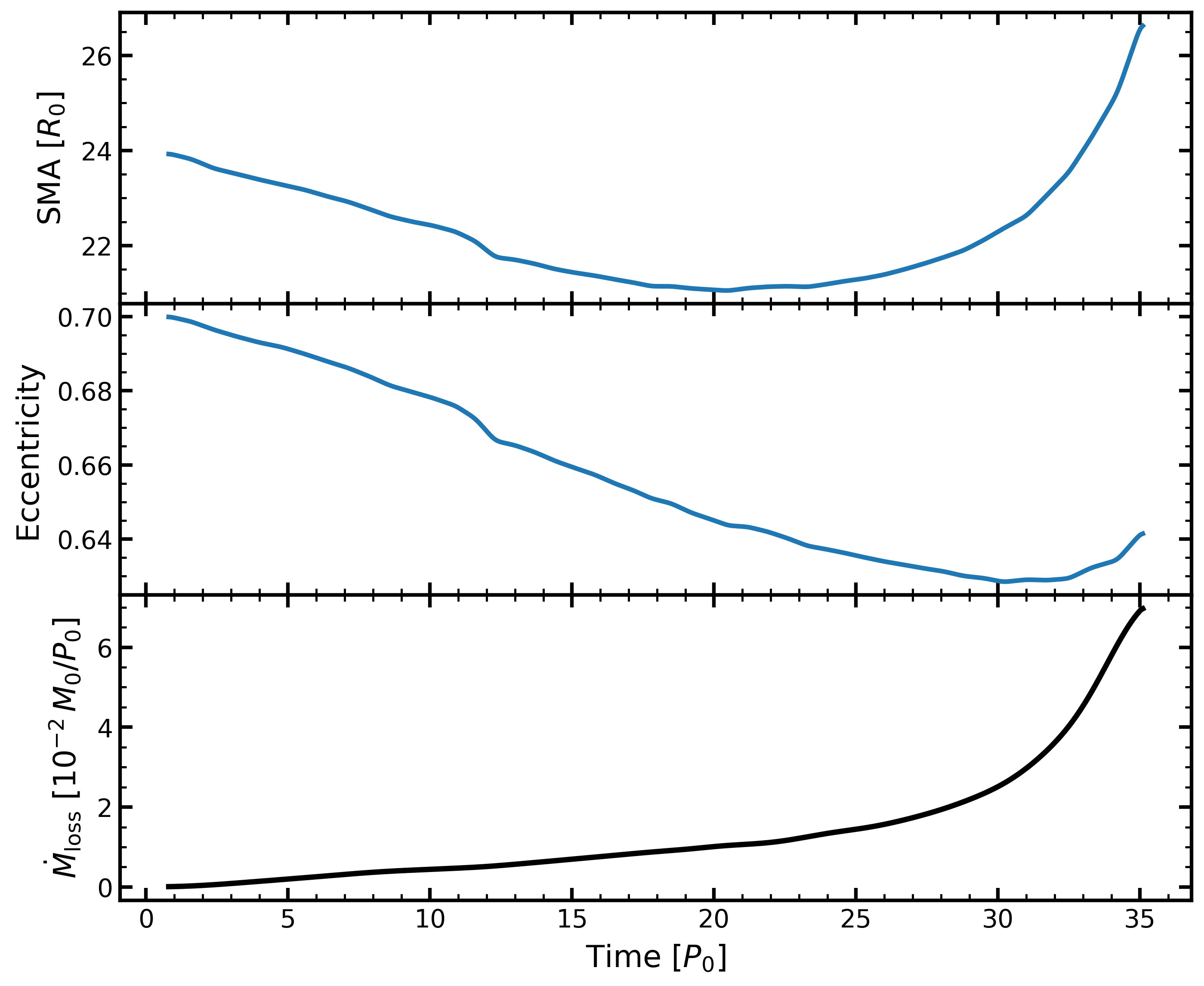}
	\caption{Results for Run E (with $b_0=3.33$ and $e_0=0.70$). The cumulative mass loss reaches about $0.8\,M_0$ by $t\simeq 35 P_0$, followed by disruption.}
	\label{fig:runE_summary}
\end{figure}

\begin{figure}[htbp]
	\centering
	\includegraphics[width=0.88\linewidth]{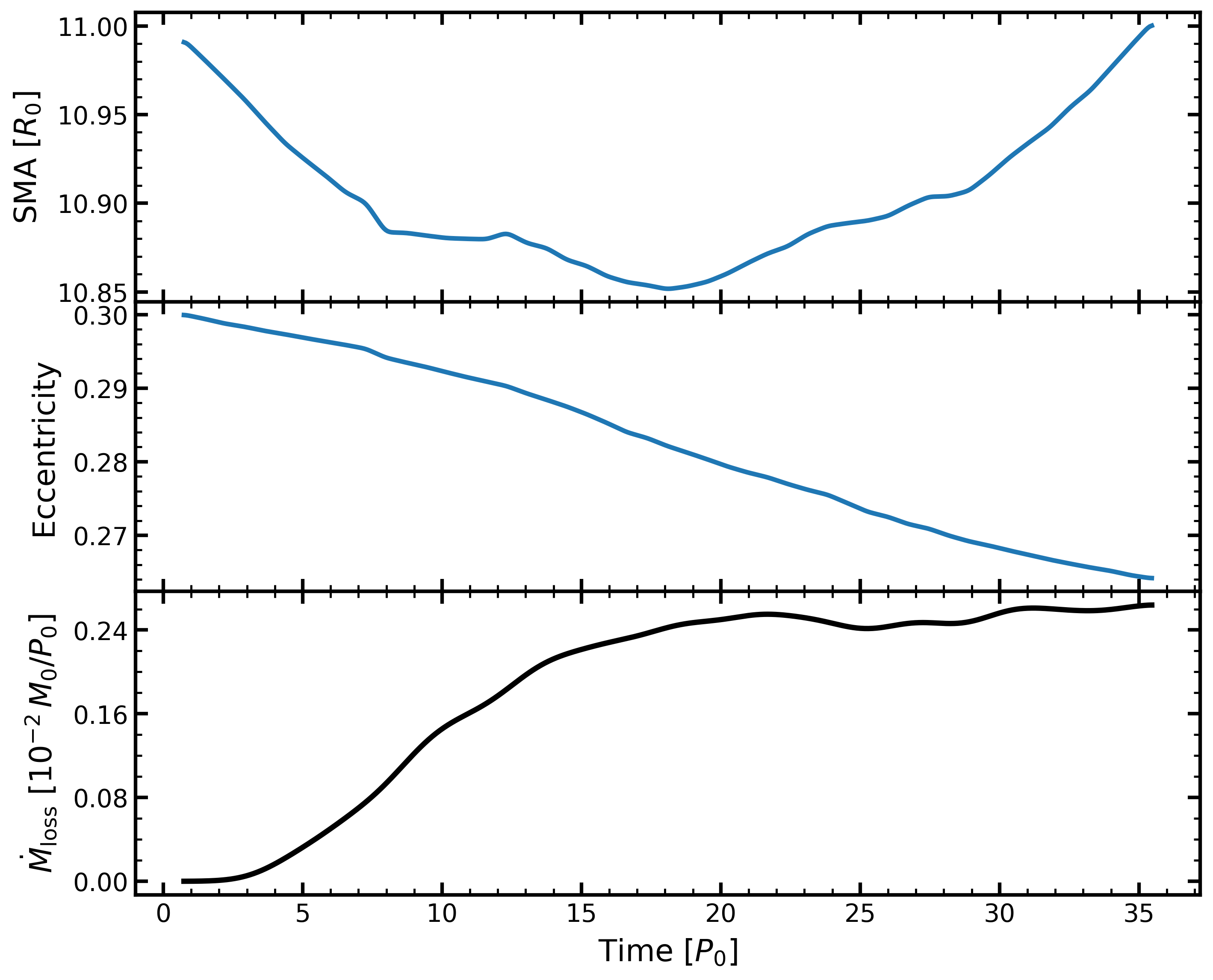}
	\caption{Results for Run F (with $b_0=3.57$ and $e_0=0.30$). The cumulative mass loss at $t = 35 P_0$ is about $0.05\,M_0$, and the binary reaches a stable mass-transfer state.}
	\label{fig:runF_summary}
\end{figure}

Run C has the same eccentricity as Runs A and B ($e_0=0.55$), but starts at the intermediate distance $b_0=3.45$ (see Fig.~\ref{fig:runC_summary}). It therefore probes the transition region between the two outcomes. Initially, the system experiences a prolonged ``deceptive'' phase ($t \lesssim 70 P_0$), during which the mass loss rate rises slowly and can resemble a self-regulated state. However, unlike Run B, the orbital widening does not fully offset the adiabatic expansion of the stellar envelope. Around $t\simeq 80 P_0$, the positive feedback loop becomes dominant, and by $t\simeq 85 P_0$, the cumulative mass loss exceeds $40\%$ of the stellar mass. This delayed runaway indicates that Run C is still on the disruptive side of the transition for $e_0=0.55$.

Runs D and E are counterparts of Run A, with the same $b_0=3.33$, but different eccentricities, $e_0=0.30$ (Run D) and $e_0=0.70$ (Run E), respectively. For Run D, the binary remains near pericenter for a larger fraction of the orbit, and the star gets disrupted earlier than in Run A (see Fig.~\ref{fig:runD_summary}). For Run E, the binary samples the strong tidal region more impulsively, but still undergoes runaway disruption (see Fig.~\ref{fig:runE_summary}). These two runs show that, at this pericenter distance, the runaway outcome is robust over the eccentricity range studied here.
The behaviors of Runs D and E further suggest that the runaway-disruption outcome found at this pericenter distance may extend to even higher eccentricities.

Run F is the low-eccentricity counterpart of Run B, with $b_0=3.57$ and $e_0=0.30$ (see Fig.~\ref{fig:runF_summary}). Despite the longer dwell time near the BH, the system does not run away. Instead, mass transfer remains regulated, similar to Run B.

Together, Runs C--F indicate that $b_0$ shapes the transition between stable eccentric mass transfer and runaway disruption. At $b_0=3.33$, the systems disrupt for the eccentricities tested here ($e_0=0.30$--$0.70$), with the lower-eccentricity case disrupting earlier. At $b_0=3.57$, however, both Run B and the low-eccentricity Run F remain in a regulated mass-transfer state.

	\section{SUMMARY AND DISCUSSION} \label{sec:conclusions}

	In this work, we have performed long-term hydrodynamical simulations using \textsc{Phantom} to investigate the fate of a Sun-like star ($1\,M_\odot$) interacting with a stellar-mass black hole ($10\,M_\odot$) on eccentric orbits. We explored initial eccentricities $e_0=0.30$, $0.55$, and $0.70$, and dimensionless pericenter distances $b_0\equiv r_{p,0}/r_{\rm tide,0}=3.33$, $3.45$, and $3.57$, following the systems for tens to over 100 $P_0$ (with $P_0$ the initial orbital period). Our main findings are as follows (see Table~\ref{tab:initial_conditions}).

	$\bullet$ Run A ($e_0=0.55$, $b_0=3.33$) demonstrates a hydrodynamical route from eccentric mass transfer to runaway stellar disruption. Once mass stripping begins, the stellar envelope expands adiabatically, increasing the Roche-lobe filling factor and driving further mass loss. This positive feedback causes superlinear growth of the cumulative mass loss and ultimately destroys the star at $t\sim 40 P_0$. The mass accretion rate onto the BH (evaluated at sink radius $500 GM_{\rm BH}/c^2$) peaks at about $5\times 10^6\,\dot{M}_{\rm Edd}$ and gradually declines after stellar disruption. Run C, at $b_0=3.45$, undergoes delayed stellar disruption.

	$\bullet$ A modest change in the initial pericenter distance can qualitatively change the outcome. Run B, with the same eccentricity ($e_0=0.55$) but $b_0=3.57$, does not undergo runaway disruption over the simulated duration ($150 P_0$). Its mass loss rate stabilizes, and orbital widening regulates the Roche-lobe filling factor and mass transfer. The BH mass accretion rate reaches about $10^5\,\dot{M}_{\rm Edd}$ at later times. Stable mass transfer is also attained for Run F, with $e_0=0.30$ and $b_0=3.57$.

	$\bullet$ At $b_0=3.33$, simulations with both lower and higher eccentricities (Runs D and E, with $e_0=0.30$ and $0.70$) lead to stellar disruption at $t\simeq 32 P_0$ and $40 P_0$, respectively, suggesting that the runaway outcome at this pericenter distance may extend to even higher eccentricities.

	The scenarios explored in this paper sit between the stable Roche-lobe overflow regime for circular binaries \citep[e.g.,][]{2025A&A...702A..61R} and the deeply unstable tidal-peeling regime for eccentric binaries \citep{2024ApJ...961..149X}. Overall, our simulations show that close, eccentric star-BH binaries can evolve along two distinct tracks: runaway disruption or self-regulated mass transfer. These two outcomes are set by the competition between stellar expansion and orbital expansion. Mass loss drives adiabatic expansion of the stellar envelope, increasing the Roche-lobe filling factor and promoting further stripping; conversely, mass loss can widen the orbit and enlarge the effective Roche lobe, reducing the filling factor. Run A represents the case where the first feedback wins and produces runaway disruption, while Run B shows that the second feedback can regulate the system. Runs C--F indicate that the transition is not a single universal pericenter threshold; it is shaped by both $b_0$ and $e_0$.

	Finally, the runaway stellar disruption (micro-TDE) represented by Run A produces a thick accretion flow and a highly super-Eddington mass supply onto the BH. Such a flow may drive powerful optically thick outflows, whose reprocessed emission could appear as a fast soft X-ray/UV transient or an apparent ULX-like source. If optical reprocessing is efficient, the timescale and luminosity may overlap with fast blue optical transients \citep{2021ApJ...911..104K,2023ApJ...949..120H}. Shocked debris and possible jet launching could add hard X-ray or radio emission \citep{2025ApJ...994L..17O,2025arXiv250922779B}. In contrast, the regulated mass-transfer branch represented by Run B should be longer lived and less explosive, possibly generating repeating, quasi-periodic flares rather than a one-off disruption \citep[e.g.,][]{2023MNRAS.524.6247L,2025arXiv250510611Y}. Future work with a denser parameter survey, radiative transfer, and magnetic fields will be needed to connect these hydrodynamical outcomes to multi-wavelength observations of electromagnetic transients.

	\software{
		PHANTOM \citep{2018PASA...35...31P}, 
		MESA \citep{2011ApJS..192....3P}, 
		yt \citep{2011ApJS..192....9T}
	}
	
	\bibliography{sample701}{}
	\bibliographystyle{aasjournal}
\end{document}